\documentclass[%
reprint,
superscriptaddress,
notitlepage,
amsmath,amssymb,
aps,
onecolumn,
floatfix,
tightenlines,
longbibliography,
11pt,
]{revtex4-1}
\usepackage{psfrag,graphicx,epsfig,color}
\usepackage{dcolumn}
\usepackage{bm}
\usepackage{natbib}
\usepackage{float}
\usepackage[usenames,dvipsnames,svgnames,table]{xcolor}
\usepackage{subfigure}

\graphicspath{{figs/}{plots/}}

\def\re    {{R_\lambda}}
\def\uu {{\mathbf{u}}}
\def\ww {{\boldsymbol{\omega}}}
\def\ew {{\hat{\ww}}}
\def\ei {{\mathbf{e}_i}}
\def\wsw {{\omega_i \omega_j s_{ij}}}

\begin{document}

\title{
Vortex stretching and enstrophy production in high Reynolds number turbulence  
}

\author{Dhawal Buaria }
\email[]{dhawal.buaria@ds.mpg.de}
\affiliation{Max Planck Institute for Dynamics and Self-Organization, 37077 G\"ottingen, Germany}
\affiliation{Department of Mechanical and Aerospace Engineering, New York University, New York, NY 11201, USA}
%

\author{Eberhard Bodenschatz}
\affiliation{Max Planck Institute for Dynamics and Self-Organization, 37077 G\"ottingen, Germany}
\affiliation{Institute for Nonlinear Dynamics, University of G\"ottingen, 37077 G\"ottingen, Germany}
\affiliation{Laboratory of Atomic and Solid-State Physics and Sibley School of Mechanical and Aerospace Engineering, Cornell University, Ithaca, New York 14853, USA}

\author{Alain Pumir}
\affiliation{Laboratoire de Physique, ENS de Lyon, Universit\'e de Lyon 1 and CNRS, 69007 Lyon, France}
\affiliation{Max Planck Institute for Dynamics and Self-Organization, 37077 G\"ottingen, Germany}

\date{\today}


\begin{abstract}

An essential ingredient of turbulent flows is the
vortex stretching mechanism, which emanates 
from the non-linear interaction of vorticity and strain-rate tensor
and leads to formation of extreme events.
We analyze the statistical correlations between vorticity 
and strain rate by using a massive database generated from very well resolved direct
numerical simulations of forced isotropic turbulence in periodic domains.
The grid resolution is up to 
$12288^3$, and the Taylor-scale
Reynolds number is in the range  $140-1300$.
In order to understand the formation and structure of extreme vorticity fluctuations,
we obtain statistics conditioned on enstrophy (vorticity-squared).
The magnitude of strain, as well as its eigenvalues,
is approximately constant when conditioned on weak enstrophy;
whereas they grow approximately as power laws for strong enstrophy,
which become steeper with increasing $\re$. 
We find that the well-known preferential alignment between vorticity and the 
intermediate eigenvector of strain tensor is even stronger
for large enstrophy, whereas
vorticity shows a tendency to be
weakly orthogonal to the 
most extensive eigenvector (for large enstrophy).
Yet the dominant contribution to the production
of large enstrophy events arises from the 
most extensive eigendirection, the more so as $\re$ increases.
Nevertheless,  the stretching in intense vorticity regions is
significantly depleted, consistent with 
the kinematic properties of weakly-curved tubes in which they are organized.
Further analysis reveals that intense enstrophy is primarily depleted
via viscous diffusion, though viscous dissipation 
is also significant. 
Implications for modeling are nominally addressed 
as appropriate.

\end{abstract}

\maketitle

\section{Introduction}

Small-scale intermittency, a hallmark of fluid turbulence, 
refers to the occurrence of sudden and intense 
fluctuations of velocity gradients \cite{Frisch95,SA97}, as routinely
reflected in long tails of their
strongly non-Gaussian probability distributions \cite{MS91,Jimenez93,Ishihara09,BPBY2019}.
Understanding such extreme events is of paramount practical interest. 
For example,
strong strain rates can dramatically 
enhance mixing of scalars or
adversely affect flame propagation in 
reacting flows \cite{Pitsch2000, sreeni04},
whereas strong vortical motions 
engender clustering of particles, facilitating cloud and rain 
formation \cite{collins97,Falkovich_2002}.
From a fundamental standpoint,
amplification of gradients is an essential
component of
the energy cascading process, leading to generation
of small scales in turbulent flows \cite{tl72,jimenez2000,carbone20}.
Thus, characterizing intermittency and 
the associated generative mechanisms is at the heart of
turbulence theory and modeling \cite{Tsi2009,Meneveau11}.

\begin{figure}
\begin{center}
\includegraphics[clip=true,width=0.47\textwidth]{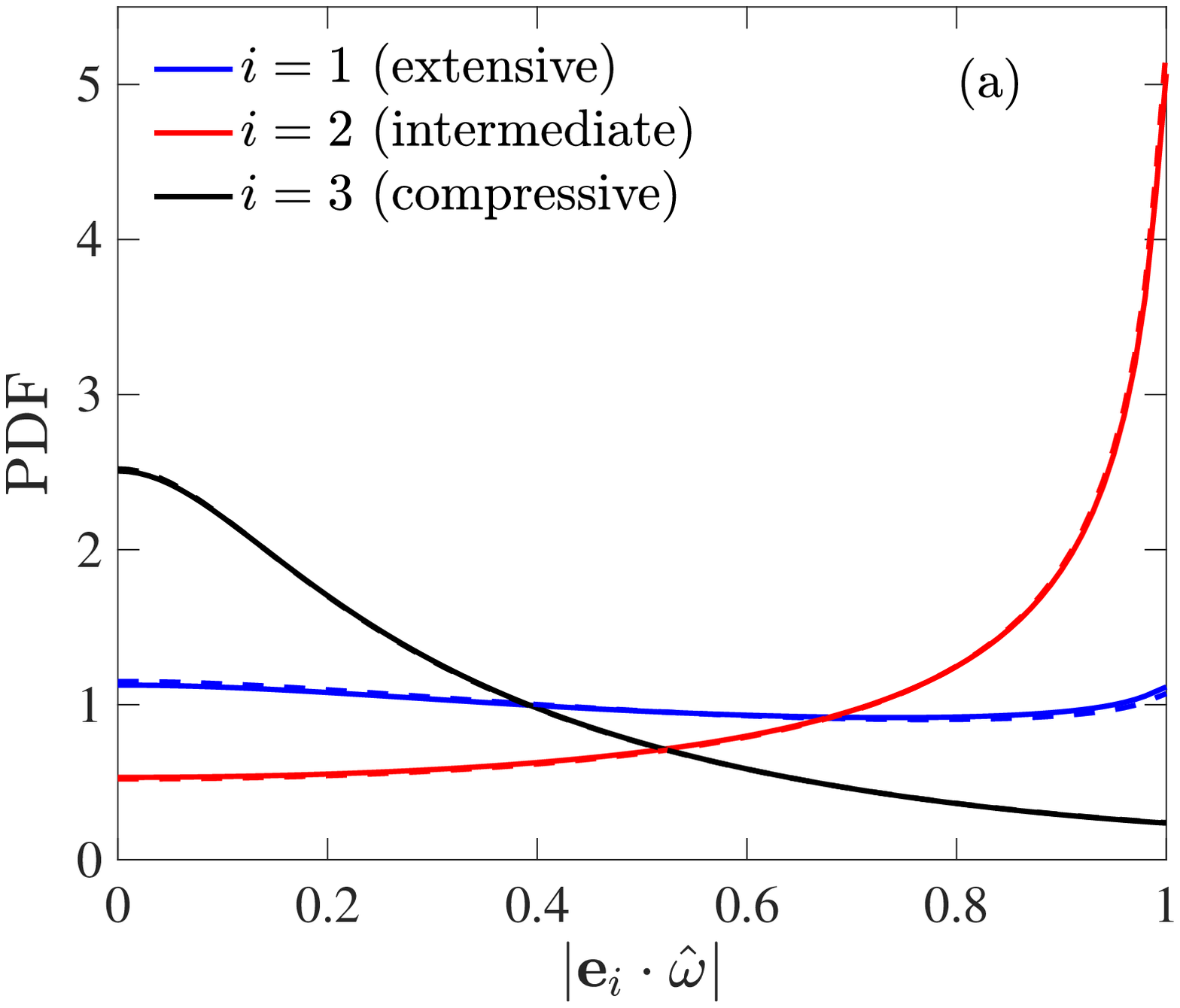} \ \ \ \ 
\includegraphics[clip=true,width=0.47\textwidth]{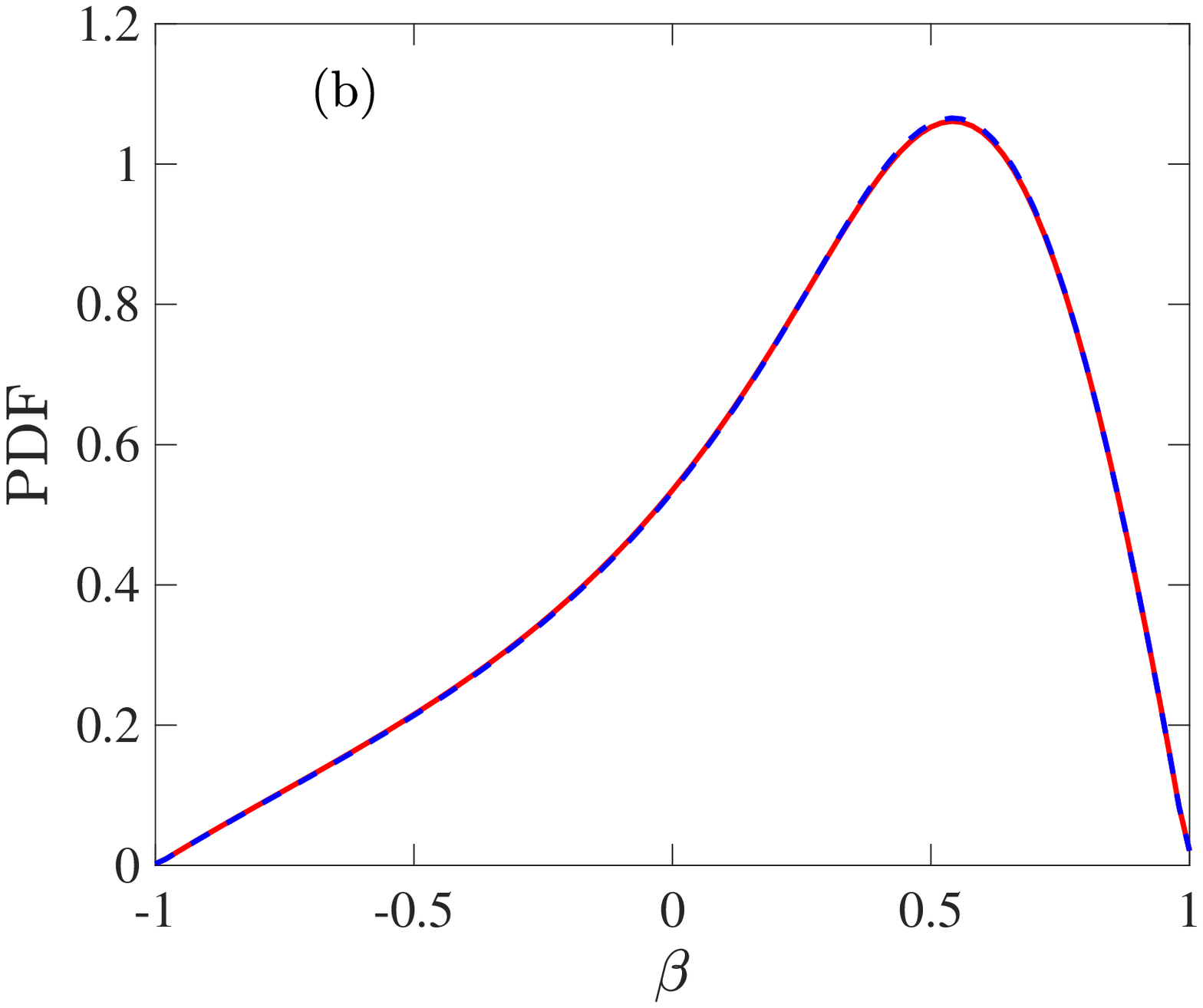}
\caption{
(a) Probability density functions (PDFs) of alignments between the vorticity unit vector 
$\hat{\ww}$, and the eigenvectors
$\mathbf{e}_i$, corresponding to the eigenvalues 
$\lambda_i$ of the strain-rate tensor 
(with $\lambda_1 \ge \lambda_2 \ge \lambda_3$).
(b) PDF of $\beta$, defined in Eq.~\eqref{eq:beta}, which 
measures the relative strength of the intermediate
eigenvalue with respect to the overall strain amplitude. 
Solid lines for Taylor-scale Reynolds number $\re=1300$ and dashed lines 
for $\re=140$.
There seems to be virtually no effect of increasing $\re$ on the PDFs.
}
\label{fig:g_align}
\end{center}
\end{figure}

A key mechanism responsible for the formation of 
such extreme events and the small-scales 
is the process of vortex stretching \cite{tl72}, 
which results from non-linear coupling between
vorticity and strain-rate tensor, respectively defined as, 
$\ww = \nabla \times \uu$, and $s_{ij} =
\frac{1}{2} ( \partial u_i/\partial x_j + \partial u_j / \partial x_i )$,
where $\uu$ is the velocity field.
This is evident from the vorticity transport equation:
\begin{equation}
\frac{D \omega_i}{D t} 
= \omega_j s_{ij} + \nu \nabla^2 \omega_i \ ,
\label{eq:vort}
\end{equation}
where 
$D/Dt = \partial_t + \uu \cdot \nabla$
is the material derivative, 
$\nu$ is kinematic viscosity and the vector 
$\omega_j s_{ij}$ gives the aforementioned
vortex stretching term. 
Numerous studies over the last 
few decades, both from experiments 
and direct numerical simulations (DNS), have revealed some robust
features of vortex stretching, which seemingly appear to be 
universal across different turbulent flows.
Notably, the vorticity vector   
preferentially
aligns with the eigenvector corresponding to the intermediate eigenvalue
of the strain tensor, which in turn is positive on average 
\cite{Ashurst87,Tsi2009,buxton10,Meneveau11,Pum16,zhou16} --
leading to 
net production of enstrophy (vorticity-squared) in
turbulent flows \cite{Betchov56}.
Fig.~\ref{fig:g_align} summarizes these findings
utilizing data from DNS of isotropic turbulence
(the details of which are discussed in Section \ref{sec:num}).
Quantitatively the results can be expected to be slightly different
for other turbulent flows, but qualitatively they behave
similarly.
Remarkably, the results also appear virtually 
independent of the Reynolds number. 


The persistent correlations between strain and 
vorticity demonstrated in Fig.~\ref{fig:g_align},
highlight some
important aspects of vortex stretching in turbulence.
Particularly, the alignment between vorticity and the intermediate
strain eigenvector suggests a depletion of stretching,
compared to what can be expected if vorticity were to align 
with the most extensive strain eigenvector.
However, the statistics shown in Fig.~\ref{fig:g_align} 
are obtained from a uniform sampling of the flow, 
and do not distinguish 
quiescent regions from those 
where extreme events reside. 
It is  well known that the extreme events
in the flow are organized in
localized vortex tubes, where
enstrophy  is orders of magnitude larger than its
global average \cite{Jimenez93,Ishihara07,BPBY2019},
whereas quiescent regions are often believed 
to be structure-less. 
Thus,  the mechanisms of vortex
stretching may differ significantly between such regions.

In this regard, statistics conditioned on the strength of vorticity
or strain offer the unique prospect of understanding
how small-scale structures are produced by the flow,
by specifically isolating
the extreme events from the moderate and the weak events.
This can be especially useful in understanding the dynamics of
vorticity amplification \cite{Tsi2009}.
In addition to providing fundamental insight,
such conditional statistics are also
very useful in turbulence modeling,
especially in PDF methods or similar approaches
\cite{pope1994,mui1996,wilczek09,johnson16}.

In this paper, our objective is to present 
a detailed fundamental investigation
of turbulence small-scale structure in light of vortex stretching
and resulting enstrophy production.
To this end, we utilize pseudo-spectral DNS of isotropic turbulence
in a periodic domain, which is the most efficient numerical tool to
study the small-scale properties of turbulence.
One important purpose of the current study 
is also to understand the effect of increasing Reynolds number.
With that in mind, we utilize a massive DNS database
with Taylor-scale Reynolds number $\re$ ranging from 140 to 1300
on grid sizes $1024^3$ to $12288^3$ respectively --
with particular attention on
having very good small-scale resolution
to resolve the extreme events accurately \cite{PK+18}.
We compute statistics related to vortex stretching,
conditioned on magnitude of vorticity.
We find that the alignment of vorticity
with intermediate strain eigenvector is strongly 
enhanced as vorticity becomes more intense.
This is qualitatively consistent with the 
notion that the most intense velocity gradient structure 
have a quasi two-dimensional (2D) structure,
for which the role of the largest strain eigenvalue in vortex
stretching is inhibited~\cite{Jimenez:1992,pumir:1990}.
However, contrary to this, we find that the
dominant contribution to enstrophy production
still comes from the most extensive eigendirection;
though the contribution from intermediate direction
is also comparable. 
Nevertheless, the stretching in the most intense vorticity regions
is significantly depleted. Further analysis shows
that viscous diffusion of enstrophy is primarily responsible for 
arresting extreme events
(and dominates over viscous dissipation of enstrophy).
We also propose improved functional forms for modeling
the statistics reported in this work.

The rest of the manuscript is organized as follows. In Section~\ref{sec:num}, we
present the database used in this work. Correlations
between strain and vorticity, in particular their dependence on the 
Reynolds number, are discussed in Section~\ref{sec:unc}.
Using conditional averages, the dependence of these correlations on
the vorticity are presented in Sections~\ref{sec:stat_vort} 
Finally, we summarize our results
in Section~\ref{sec:concl}.
Some details on numerical resolution, crucial in the study of
small scale turbulent statistics, are presented in the Appendix~\ref{app:resol}.

\section{Numerical approach and database} 
\label{sec:num}

The data used in the present work was generated via
DNS of the incompressible Navier-Stokes equations
\begin{align}
\frac{\partial u_i}{\partial t} 
+ u_j \frac{\partial u_i}{\partial x_j} =
= - \frac{1}{\rho} \frac{\partial p}{\partial x_i}  + \nu \nabla^2 u_i + f_i \ ,
\label{eq:NS} 
\end{align}
where $\mathbf{u}$ is the divergence free velocity field 
($\nabla \cdot \mathbf{u} = 0$), $p$ is the pressure, $\rho$ is the fluid density,
$\nu$ is the kinematic viscosity and $\mathbf{f}$ is the forcing term
at large scales to achieve a statistically stationary state \cite{EP88}.
We utilize a massively parallel 
implementation of Rogallo's pseudo-spectral algorithm \cite{Rogallo}
in a periodic $(2\pi)^3$ cubic domain,
with explicit second-order Runge-Kutta scheme for time integration.
The aliasing errors are controlled by a combination of
truncation and phase-shifting~\cite{PattOrs71},
resulting in highest resolved wavenumber given as
$k_{max}=\sqrt{2}N/3$, where $N$ is the number of grid points
in each direction.

\begin{table}[h]
\centering
    \begin{tabular}{cccccc}
\hline
    $\re$   & $N^3$    & $k_{max}\eta$ & $T_E/\tau_K$ & $T_{sim}$ & $N_s$  \\
\hline
    140 & $1024^3$ & 5.82 & 16.0 & 6.5$T_E$ &  24 \\
    240 & $2048^3$ & 5.70 & 30.3 & 6.0$T_E$ &  24 \\
    390 & $4096^3$ & 5.81 & 48.4 & 2.8$T_E$ &  35 \\
    650 & $8192^3$ & 5.65 & 74.4 & 2.0$T_E$ &  40 \\
   1300 & $12288^3$ & 2.95 & 147.4 & 20$\tau_K$ &  18 \\
\hline
    \end{tabular}
\caption{Simulation parameters for the DNS runs
used in the current work: 
the Taylor-scale Reynolds number ($\re$),
the number of grid points ($N^3$),
spatial resolution ($k_{max}\eta$), 
ratio of large-eddy turnover time ($T_E$)
to Kolmogorov time scale ($\tau_K$),
length of simulation ($T_{sim}$) in statistically stationary state
and the number of instantaneous snapshots ($N_s$) 
used for each run to obtain the statistics.
}
\label{tab:param}
\end{table}

The database used in the current work
is summarized in Table~\ref{tab:param}, along with the main
simulation parameters.
The runs with Taylor-scale Reynolds numbers, $\re$,
in the range $140 \le \re \le 650$ were also utilized in our 
recent work \cite{BPBY2019} and all have a very high spatial
resolution, 
$k_{max}\eta\approx6$,
or alternatively $\Delta x/\eta \approx 0.5$, where
$\Delta x=2\pi/N$ is the grid spacing and $\eta$
is the Kolmogorov length scale. This resolution should be compared
to the one used in comparable numerical investigations of turbulence
at high Reynolds numbers, which are mostly in the range 
$1 \le k_{max} \eta \le 2$ 
(or equivalently, $ 1.5 \le \Delta x/\eta \le 3$)~\cite{Ishihara09,BSY.2015}.
As established in our recent work \cite{BPBY2019}, 
such a resolution is more than adequate
to accurately resolve and study the small-scales at
the $\re$ considered, although the resolution constraints 
appear to be more severe as the Reynolds number increases.  

In addition to our previous
runs, we consider a new run at $\re=1300$ 
with a resolution of $k_{max}\eta\approx3$ (or $\Delta x/\eta\approx1$)
\cite{buaria.es,buaria_nc}.
While this resolution may not be completely sufficient at $\re=1300$
as per the strict resolution requirements proposed in \cite{BPBY2019},
the conditional statistics investigated here
are known to be less sensitive 
to resolution \cite{yeung2006}. 
(We also provide additional tests 
in the Appendix~\ref{app:resol}).
Additionally, the length of this new run is comparatively
short. 
However, the characteristic time scale of the most intense events
becomes much smaller than the Kolmogorov time scale $\tau_K$ as $\re$ 
increases~\cite{BPBY2019}, allowing for adequate 
statistical convergence 
by simply sampling more frequently in time \cite{PK+18}.

\section{Vorticity and strain correlations: unconditional statistics}
\label{sec:unc}

We begin by analyzing unconditional results on vortex
stretching, which will be useful for a better understanding of
the conditional statistics described in the next section.
The results presented here are generally consistent with previous
numerical studies at lower Reynolds numbers. By improving the
numerical resolution, and by extending the range of values of $\re$
considered, we provide additional insight, especially into
the Reynolds number dependence.

In order to quantify the intensity of vorticity
we simply utilize the enstrophy 
$\Omega=\omega_i\omega_i$.
From the vorticity transport equation in Eq.~\eqref{eq:vort}, the following transport
equation for enstrophy can be derived 
\begin{equation}
\frac{D \Omega}{D t} = 
 2 \omega_i S_{ij} \omega_j + \nu \nabla^2 \Omega 
- 2\nu \frac{\partial \omega_i}{\partial x_j} \frac{\partial \omega_i}{\partial x_j}
\label{eq:enstr}
\end{equation}
The viscous terms have been decomposed into a diffusion contribution
$\nu \nabla^2 \Omega$, 
and a purely dissipative contribution, 
$\epsilon_\omega = 2\nu ||\nabla \ww||^2$.  
In statistically stationary isotropic turbulence,
as considered here, applying averaging
to above equation gives
\begin{equation}
\langle \omega_i \omega_j s_{ij} \rangle = 
\nu \left\langle \frac{\partial \omega_i}{\partial x_j} 
\frac{\partial \omega_i}{\partial x_j}  \right\rangle 
\label{eq:enstr_bal}
\end{equation}
which gives the balance between production and dissipation of enstrophy.
The above relation is in fact also approximately valid for other turbulent flows
at sufficiently high Reynolds numbers \cite{tl72}. 

Our analysis is primarily focused around this enstrophy production term,
which is evidently affected by both the local alignment between 
strain and vorticity and their absolute magnitudes. 
This interaction between strain and vorticity is most readily 
described in 
the eigenframe of the strain tensor \cite{Ashurst87,Tsi92}.  
The strain tensor can be 
diagonalised into an orthonormal basis, with its eigenvalues given by $\lambda_i$
for $i=1,2,3$ (with $\lambda_1 \geq \lambda_2 \geq \lambda_3$)
and the corresponding
eigenvectors denoted by $\mathbf{e}_i$. Incompressibility imposes
$\lambda_1+\lambda_2+\lambda_3=0$, which in turn renders $\lambda_1$ to be always
positive (stretching) and $\lambda_3$ to be always negative (compressive).
Thus the enstrophy production term can be rewritten as
\begin{equation}
\langle \omega_i \omega_j s_{ij} \rangle = 
\langle \lambda_i (\mathbf{e}_i \cdot \omega)^2 \rangle \ ,
\label{eq:weiw}
\end{equation}
which isolates individual contribution from each eigendirection.

\begin{table}[h]
\centering
\begin{tabular}{l|c|c|c|c}
\hline
\hline
$\re$ & $\langle \lambda_i \rangle \tau_K$ &  $\langle \beta\rangle$ & $\langle \lambda_i^2\rangle \tau_K^2$ & 
$\langle (\ei \cdot \ew)^2 \rangle$ \\
\hline
    140 & 0.379 : 0.089 : -0.468 & 0.291 & 0.183 : 0.024 : 0.293  & 0.316 : 0.527 : 0:157   \\
    240 & 0.372 : 0.088 : -0.460 & 0.290 & 0.183 : 0.024 : 0.293  & 0.319 : 0.524 : 0.157   \\
    390 & 0.367 : 0.086 : -0.453 & 0.290 & 0.184 : 0.024 : 0.292  & 0:319 : 0.524 : 0.157   \\
    650 & 0.357 : 0.083 : -0.440 & 0.290 & 0.184 : 0.024 : 0.292  & 0.320 : 0.523 : 0.157   \\
   1300 & 0.345 : 0.080 : -0.425 & 0.290 & 0.184 : 0.024 : 0.292  & 0.321 : 0.522 : 0.157   \\
\hline
\end{tabular} 
\begin{tabular}{l|cc|c}
\ \\
\hline
\hline
$\re$ & $\langle \lambda_i (\ei \cdot \omega)^2\rangle \tau_K^3$ & $\langle \wsw \rangle\tau_K^3$   & $-\mathcal{S}$ \\
\hline
    140 & 0.115 : 0.101 : -0.056 & 0.160 & 0.524 \\
    240 & 0.123 : 0.106 : -0.061 & 0.168 & 0.552 \\
    390 & 0.131 : 0.110 : -0.066 & 0.175 & 0.585 \\
    650 & 0.144 : 0.118 : -0.074 & 0.188 & 0.630 \\
   1300 & 0.163 : 0.129 : -0.084 & 0.208 & 0.695 \\
\hline
    \end{tabular}
\caption{Unconditional values for various quantities in three eigendirection. Each quantity
is appropriately normalized by the Kolmogorov time scale $\tau_K$.
}
\label{tab:unc}
\end{table}

\begin{figure}
\begin{center}
\includegraphics[width=0.32\textwidth]{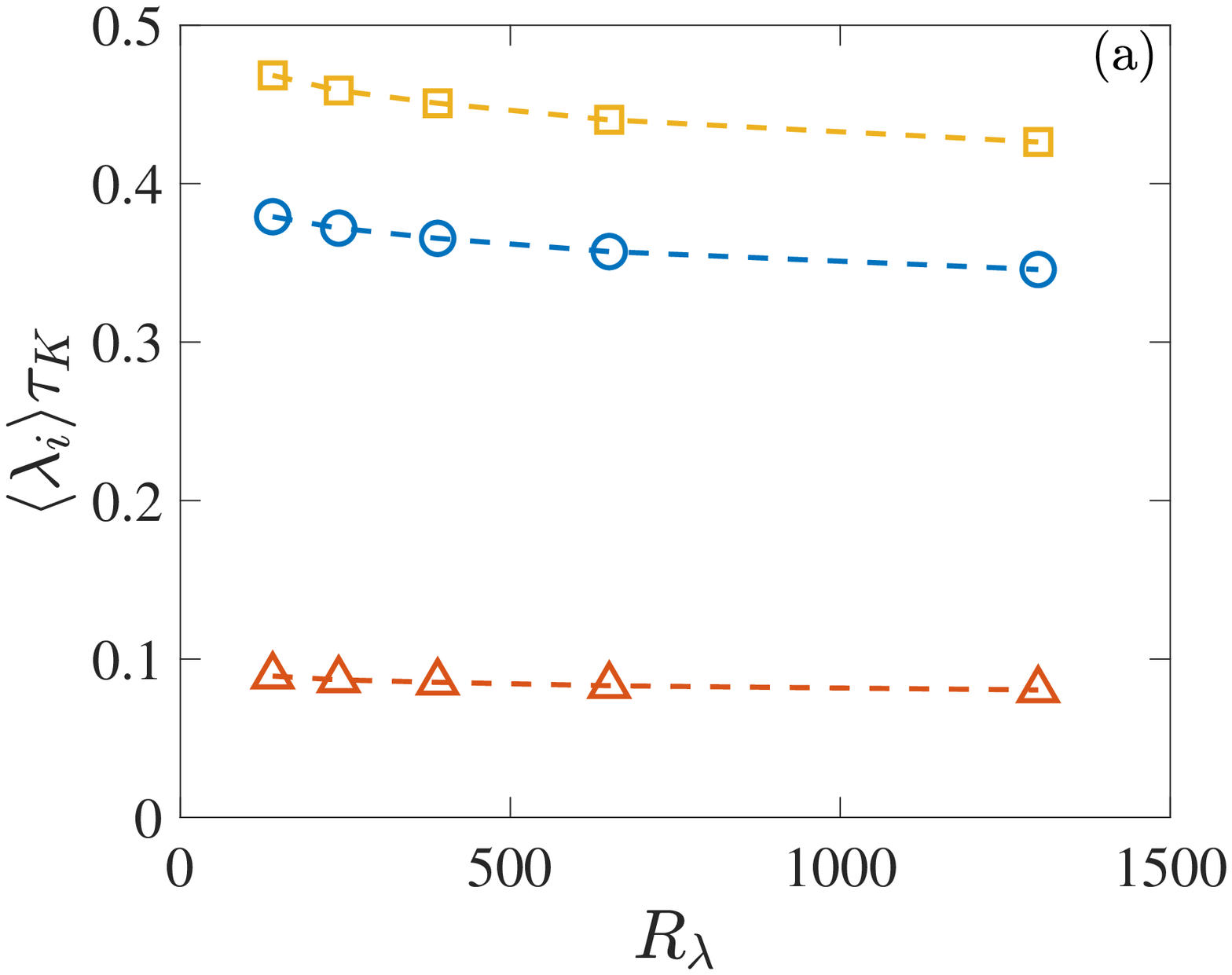} \ 
\includegraphics[width=0.31\textwidth]{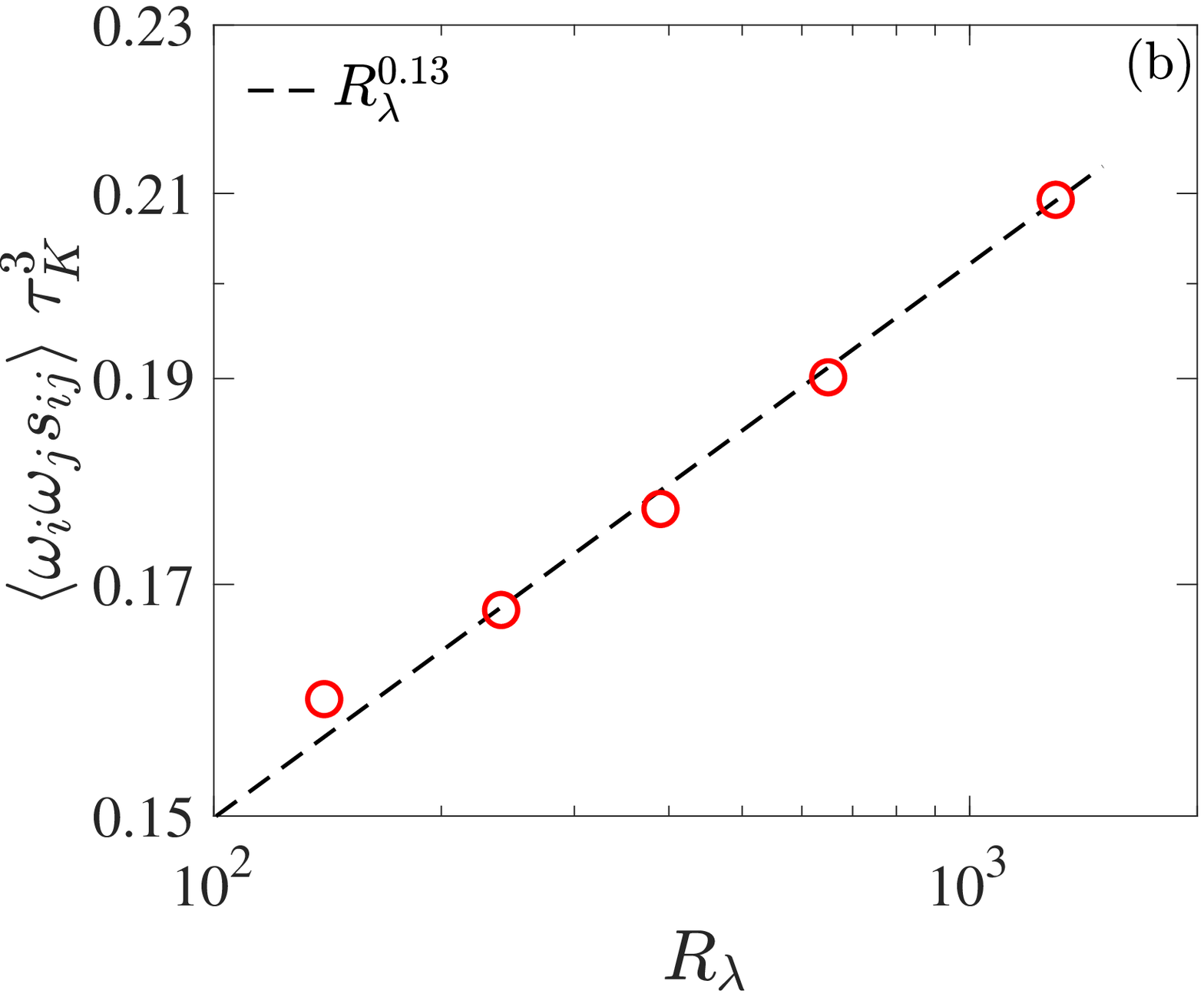} \
\includegraphics[width=0.32\textwidth]{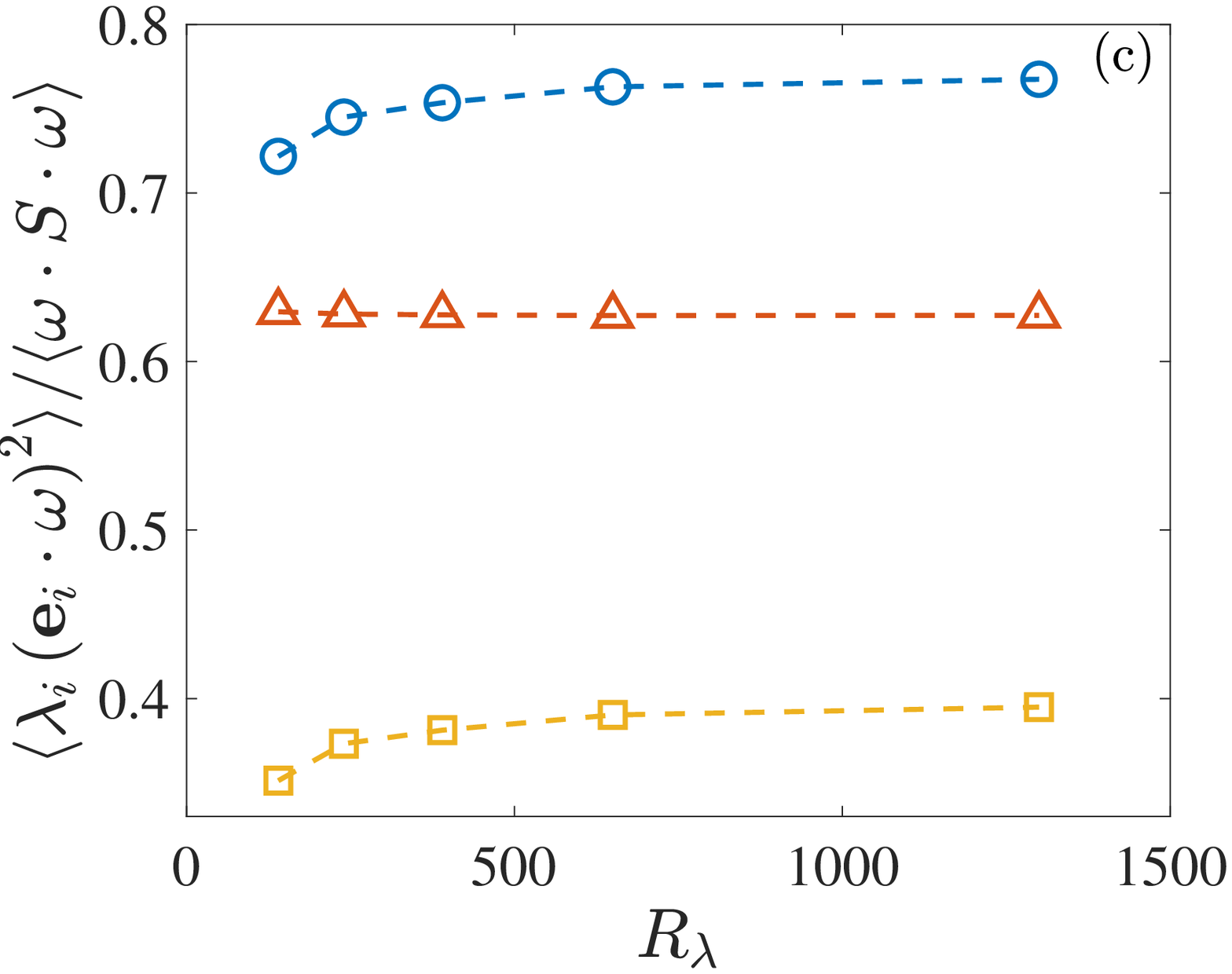} 
\caption{
Plots of various quantities from from Table~\ref{tab:unc}:
(a) eigenvalues of strain,
(b) net production of enstrophy, and
(c) relative contribution from each eigendirection to
the  net enstrophy production.
In (a) and (b), blue circles, red triangle and yellow squares
represent $i=1,2$ and $3$ respectively. For $i=3$, the 
negative of the value is shown in both (a) and (c).
}
\label{fig:unc}
\end{center}
\end{figure}

Various individual contributions in each eigendirection are computed
and listed in Table~\ref{tab:unc} and also plotted in 
Fig.~\ref{fig:unc}.
All (dimensional) quantities are non-dimensionalized appropriately
by the Kolmogorov time scale $\tau_K$, 
defined as
\begin{align}
\tau_K = \left( \nu/ \langle \epsilon \rangle \right)^{1/2} \ , 
\end{align}
where $\epsilon  = 2\nu s_{ij} s_{ij}$ is the energy dissipation rate. 
Note that in homogeneous turbulence $\langle \Omega \rangle 
= \langle \epsilon \rangle /\nu$ and hence 
$\tau_K^2 = 1/\langle \Omega \rangle$. 
We first consider the mean of each eigenvalue as shown first in the table
and plotted in Fig.~\ref{fig:unc}a.
Note that they sum up to zero due to incompressibility. 
Interestingly, it appears the contribution in each 
direction slowly decreases with increasing $\re$,
however, the ratio 
$\langle \lambda_1 \rangle / \langle \lambda_2 \rangle$ is
approximately constant (at about 4.3). Alternatively, the quantity
$\beta$ defined as 
\begin{align}
\beta = \frac{\sqrt{6}\lambda_2}{(\lambda_1^2 + \lambda_2^2 + \lambda_3^2)^{1/2}}
\label{eq:beta}
\end{align}
which measures the relative amplitude of $\lambda_2$, is 
also
independent of $\re$.
This is indeed to be expected, based on Fig.~\ref{fig:g_align},
where we saw that the PDF of $\beta$ itself does not
change with $\re$.
Although not shown, we also observe that the PDF of
$\lambda_2/\lambda_1$ is also independent 
of $\re$ (and qualitatively similar to that of $\beta$).
Additionally, their PDFs suggest 
that the most probable value of 
$\lambda_1 : \lambda_2 : \lambda_3$ 
is approximately $2.85:1:-3.85$, which is in good agreement 
with earlier works \cite{Ashurst87,Tsi92}.

Likewise, the mean-square values of individual
eigenvalues (the next quantity in Table~\ref{tab:unc}),
once scaled with $\tau_K^2$, are essentially independent of $\re$.
Note by definition, they sum up to $1/2$, since $\lambda_i \lambda_i = s_{ij} s_{ij}$.
Similarly, the mean-square of alignment cosine between vorticity and
eigenvectors of strain -- defined as 
$\langle (\mathbf{e}_i \cdot \hat{\ww})^2\rangle $, where
$\hat{\ww}$ is the unit vector along vorticity -- 
also do not reveal any dependence on $\re$.
This is again consistent with Fig.~\ref{fig:g_align}a, where
the PDFs of these cosines were seen to be practically identical at
the lowest and highest $\re$ in this work.  
Note that the square of cosines is considered because they have the 
nice property of summing up to
unity for all three directions (in addition to being
bounded between 0 and 1 for each individual direction
for the case of perfect orthogonal and parallel alignment respectively).

Next in Table~\ref{tab:unc}, we consider the individual contribution to net enstrophy 
production through the terms $\langle \lambda_i (\mathbf{e}_i \cdot \omega)^2 \rangle$. 
It follows that the contribution from the
first direction is always positive
and that from the third direction is always negative. We find that  
$\langle \lambda_1 ( \mathbf{e}_1 \cdot \ww)^2 \rangle$
is approximately twice as large as
$- \langle \lambda_3 ( \mathbf{e}_3 \cdot \ww)^2 \rangle$.
Additionally, the intermediate direction is strongly positive
resulting in net positive enstrophy production.
Once non-dimensionalized, we find that the absolute
values of the three components
$\langle \lambda_i ( \mathbf{e}_i \cdot \omega)^2 \rangle \tau_K^3$
all increase with $\re$.
Importantly, despite the strong alignment with intermediate eigenvector,
the net contribution to enstrophy production from the first eigendirection
is always larger than the second. In fact, the difference
between the contributions of the first and second strain eigendirections
increases with $\re$.

To further clarify, the individual
contribution in each eigendirection as a fraction of total production
is shown in Fig.~\ref{fig:unc}c 
(absolute value is shown 
for comparison). Interestingly, we find that the 
relative contribution from the intermediate direction is 
virtually constant with increasing $\re$, whereas the 
contributions from the first and third directions
both increase in the same manner (though they are opposite in signs). 
However, it appears that this increase is prominent only
at smaller $\re$ and an asymptotic state is 
possibly reached at high $\re$, such that the relative contribution
from each eigendirection is constant, and 
approximately in the ratio $0.78:0.62:-0.40$.

Whereas our results summarized in Table~\ref{tab:unc} and Fig.~\ref{fig:unc}
agree qualitatively with those of earlier studies, 
we notice a significant quantitative difference compared to
the results of \cite{Tsi92,luethi:2005,Tsi2009}, especially
concerning the individual eigenvalues and their contribution
to net enstrophy production. On the other hand, results 
of \cite{zhou16} at $\re\sim100$ based on simulations of grid turbulence, 
appear to be in reasonable agreement with our results
at $\re=140$. This suggests that the disagreement with 
the results of \cite{Tsi92,luethi:2005,Tsi2009}
is likely due to practical difficulties in experimentally 
measuring the velocity gradient tensor 
with sufficient accuracy \cite{Tsi2009,Wallace10}.

Finally, the increase in net enstrophy production 
with $\re$, documented in Fig.~\ref{fig:unc}b, can be readily explained 
utilizing the well-known relation:
\begin{align}
\langle \omega_i \omega_j s_{ij} \rangle \tau_K^3 =  -\frac{7\sqrt{15}}{90} \mathcal{S} \approx -0.3 \ \mathcal{S}
\label{eq:skewness}
\end{align} 
where $\mathcal{S}$ is the skewness of the longitudinal 
velocity gradients \cite{Betchov56,kerr85}.
It is also well known that due a forward cascade of energy in 
(three-dimensional) turbulent flows, $\mathcal{S}$ is negative and its magnitude 
slowly increases with $\re$,
approximately as a power law \cite{Ishihara07}.
The final column of Table~\ref{tab:unc} lists the values of $-\mathcal{S}$
for different $\re$, which are in excellent agreement with Eq.~\eqref{eq:skewness}. 
On a similar note, 
given the far improved small-scale resolution utilized in current work, 
our values of $\mathcal{S}$ suggest a minor correction to the power law fit
reported in \cite{Ishihara07}, with the new fit given as 
\begin{align}
-\mathcal{S} \sim (0.27\pm0.03) \ \re^{(0.13\pm0.02)} \ . 
\end{align}
This approximate power law also applies for the net enstrophy production
(and hence net enstrophy dissipation through Eq.~\eqref{eq:enstr_bal}),
albeit with a different prefactor.

\section{Statistics conditioned on vorticity/enstrophy}
\label{sec:stat_vort}

The unconditional correlations of 
strain and vorticity reviewed in the previous section
revealed some very robust features. 
In this section, to 
specifically understand the formation of extreme events, 
we condition the correlations
on the magnitude of vorticity.
More precisely, we use $\Omega \tau_K^2$ as 
the conditioning variable, which is simply the enstrophy
divided by its mean value, i.e., $\Omega/\langle\Omega\rangle$.
Since the distribution of $\Omega\tau_K^2$ varies
over a very large range of values, the conditioning is performed
using logarithmically spaced bins, with 8 bins
per decade (same as in \cite{BPBY2019}). 
To facilitate comparisons between runs at different $\re$, the
investigated conditional averages are also non-dimensionalized by  
Kolmogorov scales (as necessary).

\subsection{Alignment}
\label{subsec:align}

\begin{figure}
\begin{center}
\includegraphics[width=0.47\textwidth]{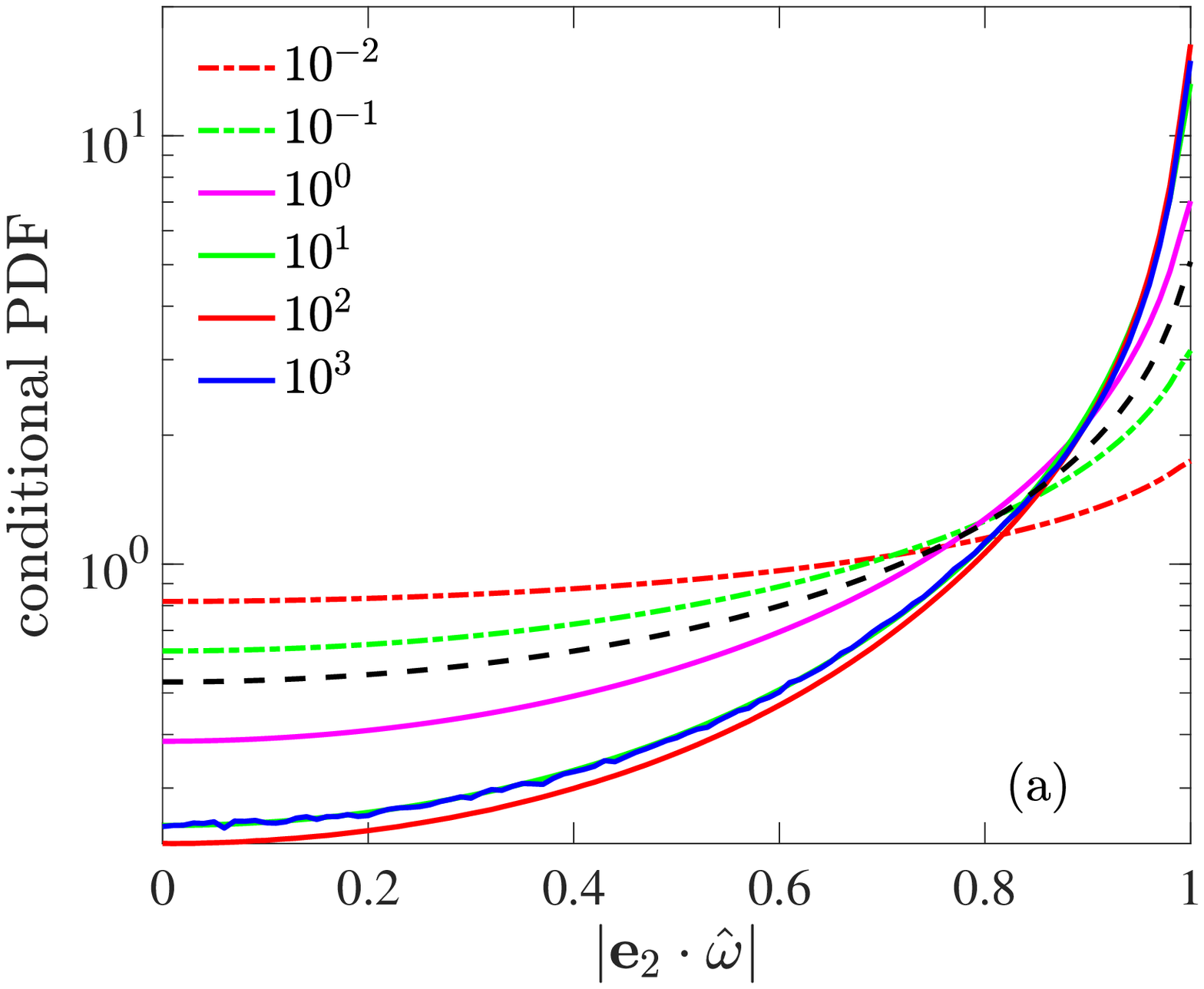} \ \ \ \
\includegraphics[width=0.47\textwidth]{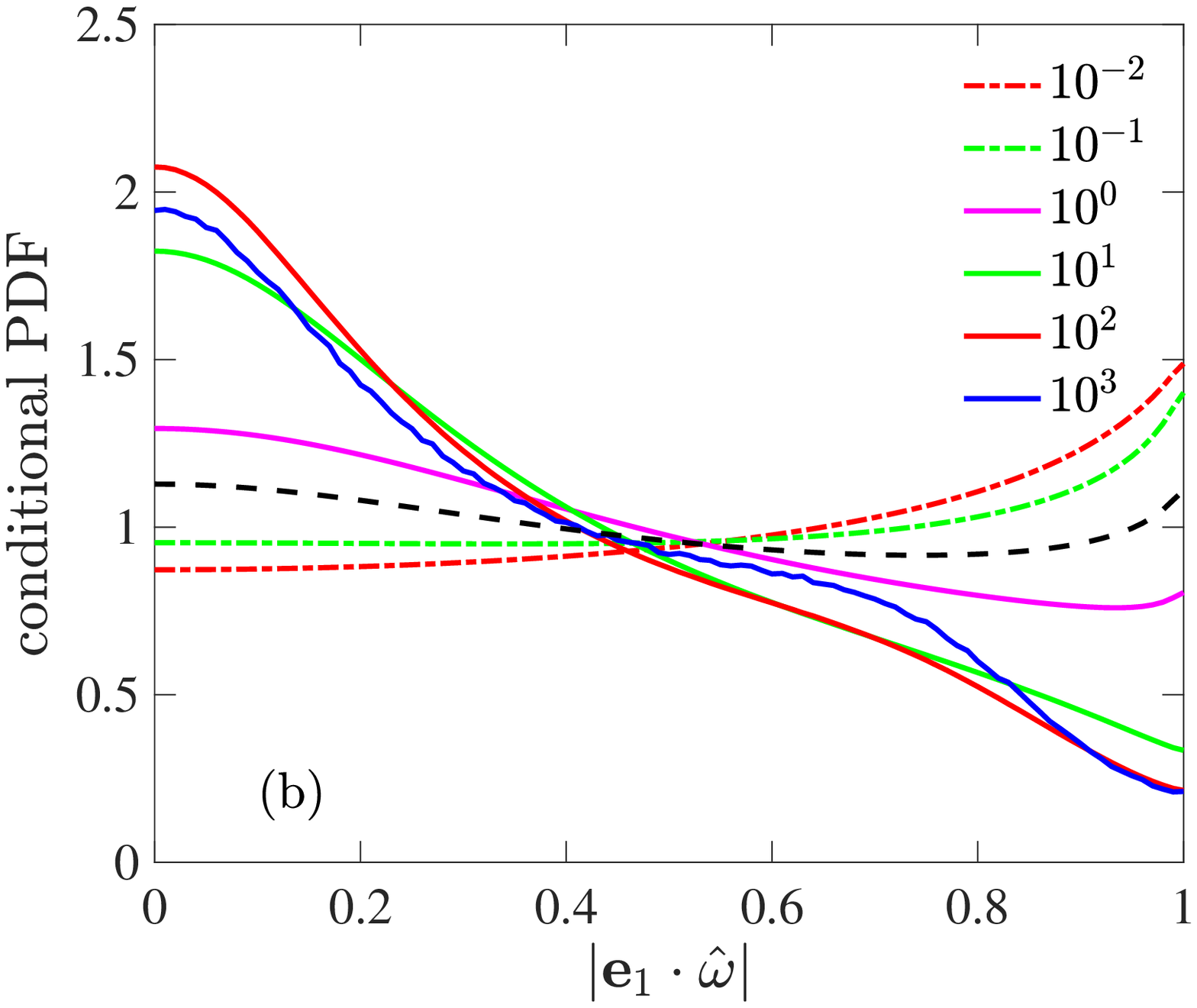} \\ 
\includegraphics[width=0.47\textwidth]{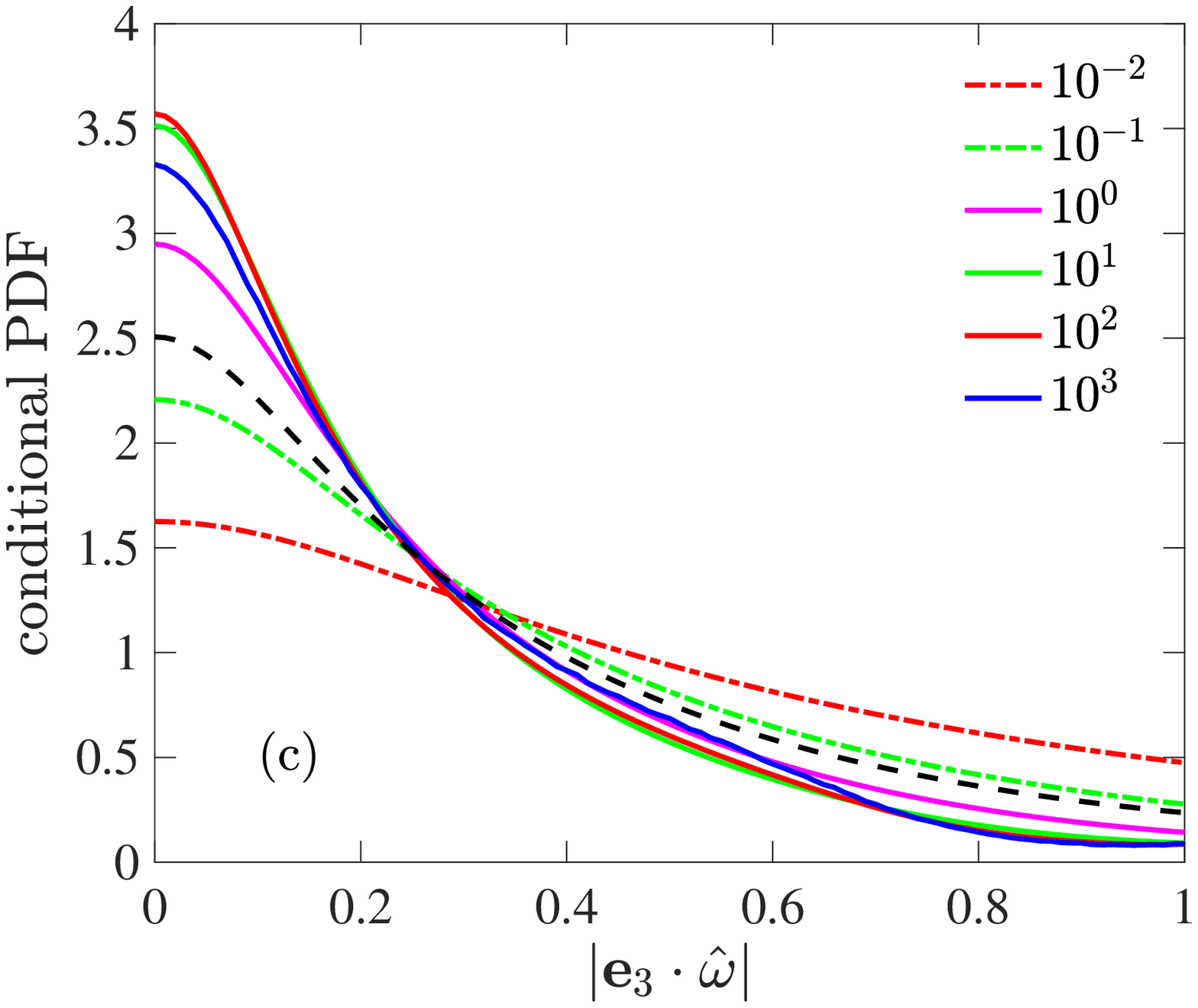} \ \ \ \
\includegraphics[width=0.47\textwidth]{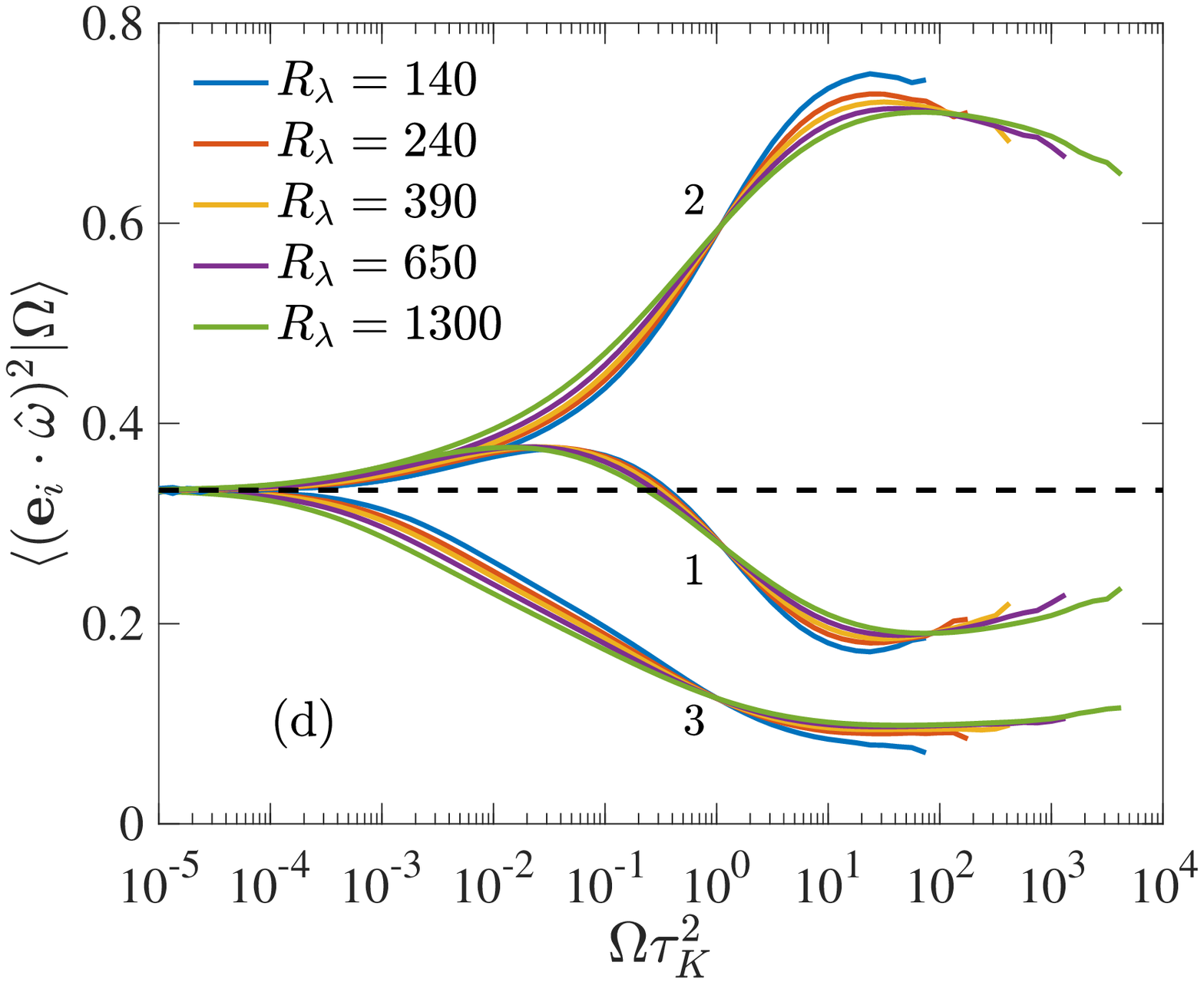} 
\caption{
Conditional probability density functions (PDFs) of 
alignment cosine between vorticity unit vector 
$\hat{\ww}$ and strain eigenvectors $\mathbf{e}_i$ 
for (a) $i=2$, (b) $i=1$, and (c) $i=3$,
at $\re=1300$ and various conditioning values 
of $\Omega\tau_K^2$ shown in legend. The black dashed
lines in each panel correspond to the marginal PDF
shown earlier in Fig.~\ref{fig:g_align}a. 
(d) The conditional second moment is shown
for each eigendirection
as a function of $\Omega\tau_K^2$ for various $\re$.
}
\label{fig:align}
\end{center}
\end{figure}

We first investigate the conditional alignment properties between vorticity 
and the eigenvectors of strain tensor by considering the
conditional PDFs of the directional cosines $\mathbf{e}_i \cdot \hat{\ww}$.
Similar to Fig.~\ref{fig:g_align}a, the absolute magnitude of the
cosine is used, since its sign
is immaterial for enstrophy production.
The conditional PDFs corresponding to the intermediate eigenvector, 
i.e., $i=2$,
are shown in Fig.~\ref{fig:align}a,
for the highest $\re$ $(=1300)$ in this work and
for the range $10^{-2} \le \Omega \tau_K^2 \le 10^3$;
for comparison the marginal PDF 
(from Fig.~\ref{fig:g_align}a)
is shown as a black dashed line.
For small values of 
$\Omega \tau_K^2$,
the preferential alignment between $\mathbf{e}_2$ and 
$\hat{\ww}$ is reduced, and PDF of $\mathbf{e}_2 \cdot \hat{\ww}$ 
approaches a uniform distribution.
On the other hand, for large values of $\Omega \tau_K^2$,
the alignment between 
$\mathbf{e}_2$ and $ \hat{\ww}$ becomes much stronger, with the
PDF of 
$\mathbf{e}_2 \cdot \hat{\ww}$. 
becoming increasingly concentrated around  $1$.
The alignment appears to be almost independent of $\Omega \tau_K^2$
for $\Omega \tau_K^2 \gtrsim 100$,
suggesting an asymptotic state for
the most extreme events.

The conditional PDFs for $i=1$ and $3$ 
are shown in Fig.~\ref{fig:align}b and c
respectively. The marginal PDFs shown in black dashed lines 
(again from Fig.~\ref{fig:g_align}a), 
did not reveal any 
preferential alignment between vorticity and $\mathbf{e}_1$, with
its distribution close to uniform; 
whereas
vorticity was preferentially orthogonal to $\mathbf{e}_3$.
The conditional PDFs of $\hat{\ww} \cdot \mathbf{e}_1$ on 
vorticity reveals a very weak preferential 
alignment between $\mathbf{e}_1$ and vorticity at small values
of $\Omega \tau_K^2$. However, this alignment disappears when
$\Omega \tau_K^2  \gtrsim 1$. For $\Omega \tau_K^2 \gtrsim 10$,
$\hat{\ww}$ and $\mathbf{e}_1$ appear to be preferentially
orthogonal to each other (albeit very weakly).
On the other hand, the conditional PDFs 
of $\hat{\ww} \cdot \mathbf{e}_3$, suggest that
vorticity is always preferentially orthogonal
to $\mathbf{e}_3$, except at small conditioning values,
where the PDF approaches an uniform distribution (similar to 
those in other directions).
To summarize, as the value of $\Omega \tau_K^2$ becomes larger,
both $\mathbf{e}_1$ and $\mathbf{e}_3$ appear to
become increasingly orthogonal to vorticity,
with the effect being strongly pronounced for $\mathbf{e}_3$. 
Simultaneously, the alignment of vorticity with $\mathbf{e}_2$
becomes extremely strong at
large $\Omega \tau_K^2$.

The above result can be concisely illustrated
by considering the second moment of the conditional 
PDFs, i.e., 
$\langle (\mathbf{e}_i \cdot \hat{\ww})^2 | \Omega \rangle$
and plotting it as a function of $\Omega \tau_K^2$. 
Since the directional cosine is bounded between 0 and 1,
the conditional expectation is also bounded between
0 and 1.
(with a value of $1/3$ for a uniformly 
distributed PDF).
In addition, the second moment has the nice property
that the contributions in three directions adds up to unity,
i.e., 
$\sum_{i=1}^3 \langle ( \mathbf{e}_i \cdot \hat{\ww})^2 | \Omega \rangle = 1$,
for any value of $\Omega$.
Such a representation also allows us to systematically
investigate the $\re$ dependence, as evident from
Fig.~\ref{fig:align}d, which shows the conditional expectation
$\langle (\mathbf{e}_i \cdot \hat{\ww})^2 | \Omega \rangle$
for $i=1,2,3$ and various $\re$.
The behavior of the conditional expectations
in Fig.~\ref{fig:align}d, is completely consistent
with that of conditional PDFs in Fig.~\ref{fig:align}a-c,
and clearly  demonstrates the 
incipient plateau-like behavior
for intense vorticity.
Interestingly,
from this figure, it is also evident
that the curves show a very weak 
dependence on $\re$.

Overall, the above results
point to very different statistical properties
of the strain field in regions of intense vorticity, compared to those
in regions of moderate or weak enstrophy.
Manifestly, the weakest events 
are `structure-less', with all alignments being equally likely. 
On the other hand, intense enstrophy regions  appear
to have a very specific structure, 
with the conditional alignments
(as seen from Fig.~\ref{fig:align}d)
approximately in the ratio $0.2:0.7:0.1$.
Since the most intense vorticity  events are
typically found to be arranged in
tubes with weak curvatures,
the above result appears consistent with an effectively
2D structure -- where
the most extensive and compressive eigenvectors
lie in the equatorial plane, 
and the intermediate eigenvector is along
the tube axis.
In fact, such an alignment effect was 
also established in \cite{Jimenez:1992},
and shown to result from kinematic constraints.
A similar behavior is also observed 
in the interaction between vortex structures in the Euler
equations, where the formation of very intense structure, shaped as
2D vortex sheets is often observed, see e.g. \cite{pumir:1990}.
For such structures, the largest and weakest strain eigenvectors are
perpendicular to vorticity, and amplification is due to the weak
intermediate eigenvector.
This results in an effective reduction of the
nonlinearity~\cite{Tsi2009}.

\subsection{Strain and eigenvalues}
\label{subsec:strain}

\begin{figure}
\begin{center}
\includegraphics[width=0.47\textwidth]{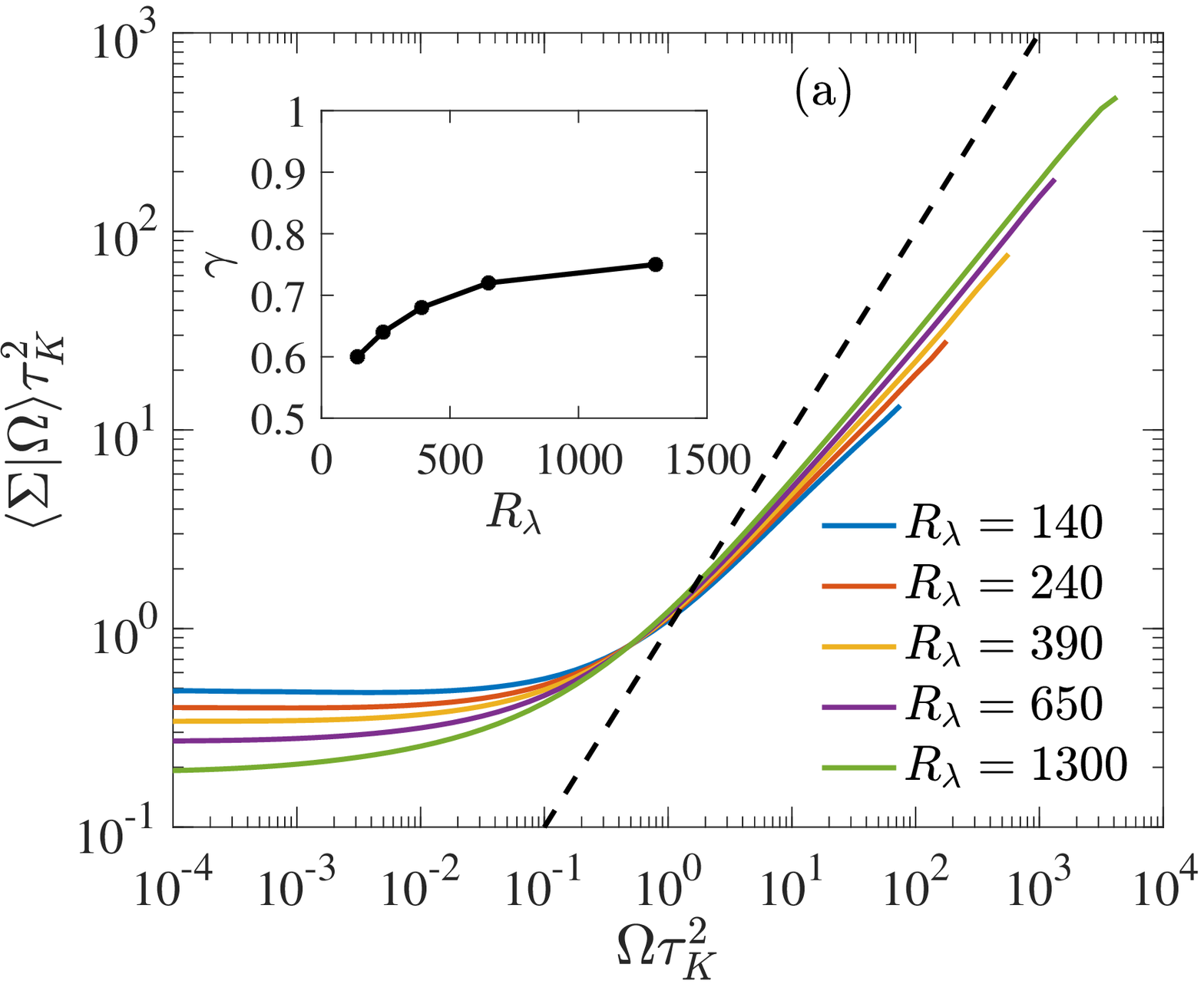} \ \ \ \
\includegraphics[width=0.47\textwidth]{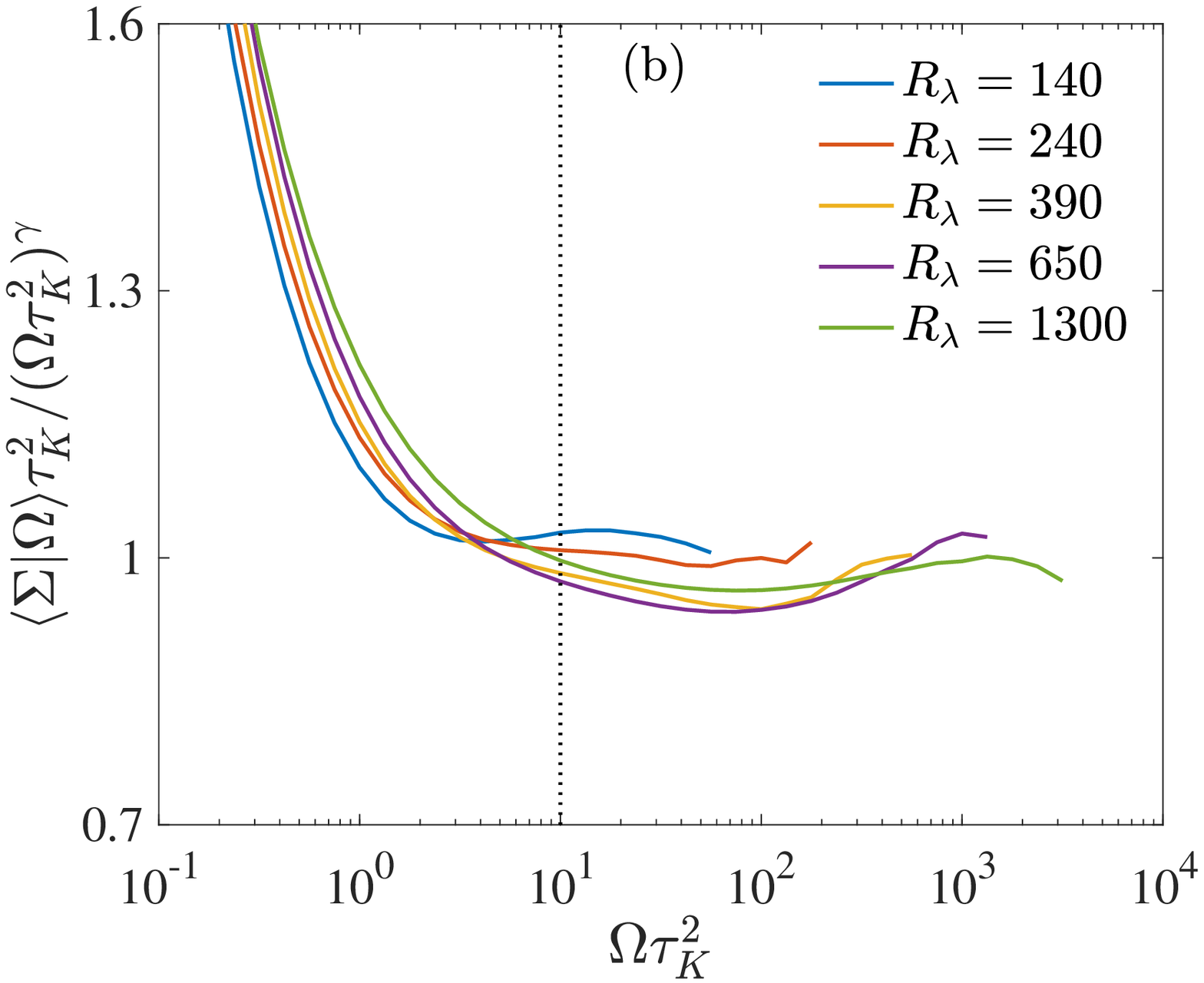} 
\caption{
(a) Conditional expectation of strain normalized by Kolmogorov 
time scale at different $\re$. The inset shows
$\gamma$ as a function of $\re$, for a power law fit of the form
$\langle \Sigma|\Omega \rangle \sim \Omega^\gamma$ in 
the region $\Omega \tau_K^2 \gtrsim 10$.
The dashed line corresponds to a power law of $\Omega^1$.
(b) Compensated conditional expectation
demonstrating the 
the robustness of the power law fit.
}
\label{fig:sigma}
\end{center}
\end{figure}

In addition to the alignments of vorticity and strain
eigenvectors, the magnitude of strain
and its eigenvalues are of obvious importance
in determining net enstrophy production.
We characterize the magnitude of the strain tensor 
by $\Sigma=2s_{ij}s_{ij}$,
which is simply the dissipation rate divided by
viscosity, i.e., $\Sigma = \epsilon/\nu$. 
In homogeneous turbulence, as considered here,
we have 
$\langle \Sigma \rangle = \langle \Omega\rangle = 1/\tau_K^2$,
allowing us to compare the extreme values of both 
$\Sigma$ and $\Omega$ with respect to their same mean value.

The conditional expectation $\langle \Sigma|\Omega \rangle$ is
shown in Fig.~\ref{fig:sigma}a.
Similar results were presented in \cite{BPBY2019},
albeit without the data at $\re = 1300$.
As noted in \cite{BPBY2019}, the conditional expectation
for smaller values of $\Omega$ is approximately constant,
suggesting vorticity and strain are uncorrelated.
Thereafter, as $\Omega$ increases, the expectation 
approximately increases as a power law,
$\langle \Sigma | \Omega \rangle \sim \Omega^{\gamma}$,
with $\gamma<1$ -- implying that strain acting on intense vortices
is comparatively weaker than the vorticity. 
The exponent $\gamma$, shown in the inset of Fig.~\ref{fig:sigma}a,
increase very weakly with $\re$,
suggesting that $\gamma=1$ would be realized as $\re\to\infty$.
Assuming a simple sigmoid or power law dependence
of $\gamma$ on $\re$ suggested that $\gamma\approx1$ would only
be realized for extremely large $\re$
(which are practically
unrealizable in both experiments and numerical simulations). 
Overall the new data at $\re = 1300$ appear
to be consistent with conclusions drawn in
\cite{BPBY2019} (which utilized data at $\re \le 650$).
In this regard, we make a note that the power law representation
$\langle \Sigma | \Omega \rangle \sim \Omega^{\gamma}$,
is an empirical observation, but nonetheless
very useful from a modeling perspective.
This power law behavior appears to be very robust,
as indicated in Fig.~\ref{fig:sigma}b, which shows the
conditional expectations compensated by $\Omega^\gamma$
-- with curves for all $\re$ showing a robust plateau 
(with variations of less than $5\%$).

\begin{figure}
\begin{center}
\includegraphics[width=0.32\textwidth]{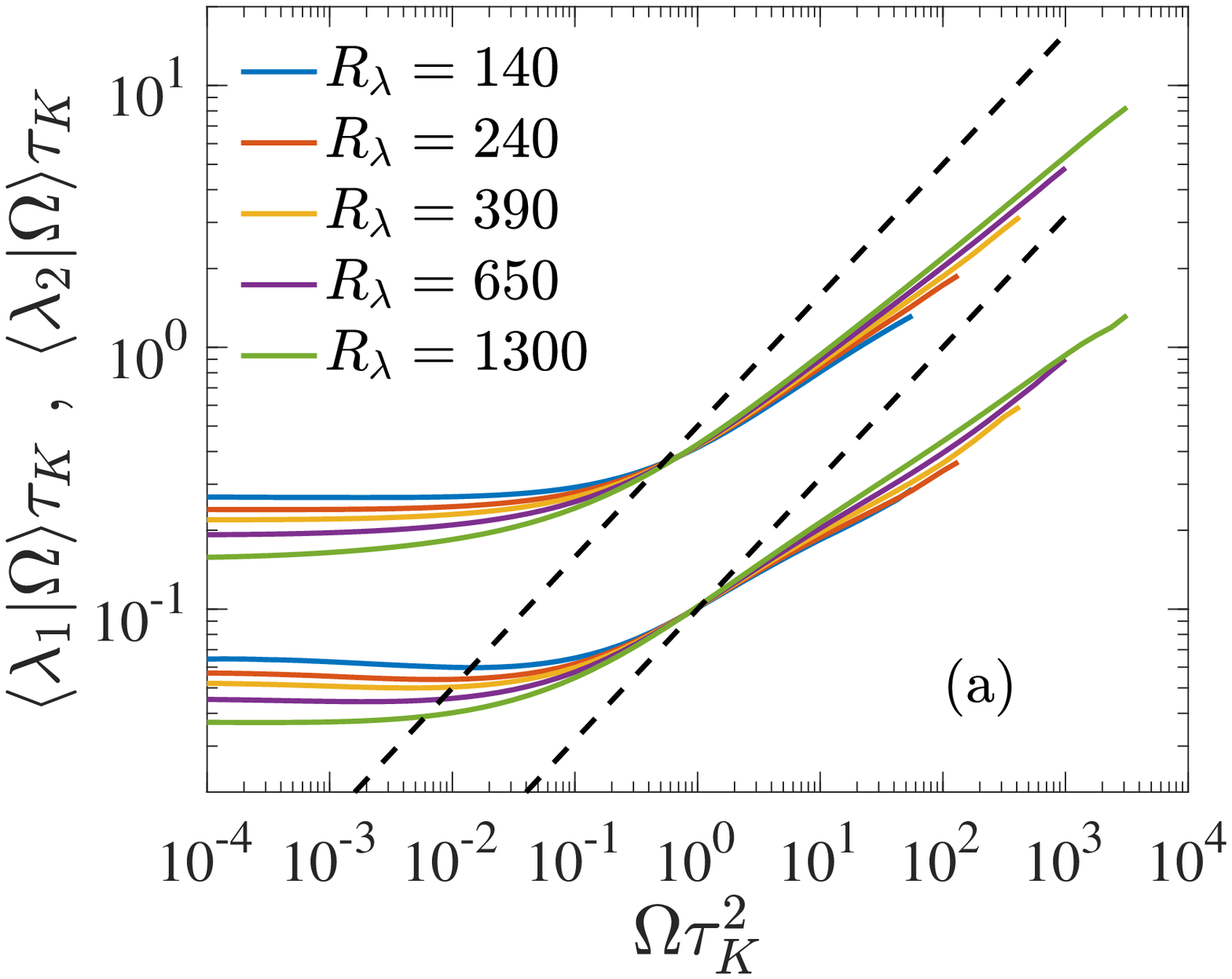} \
\includegraphics[width=0.32\textwidth]{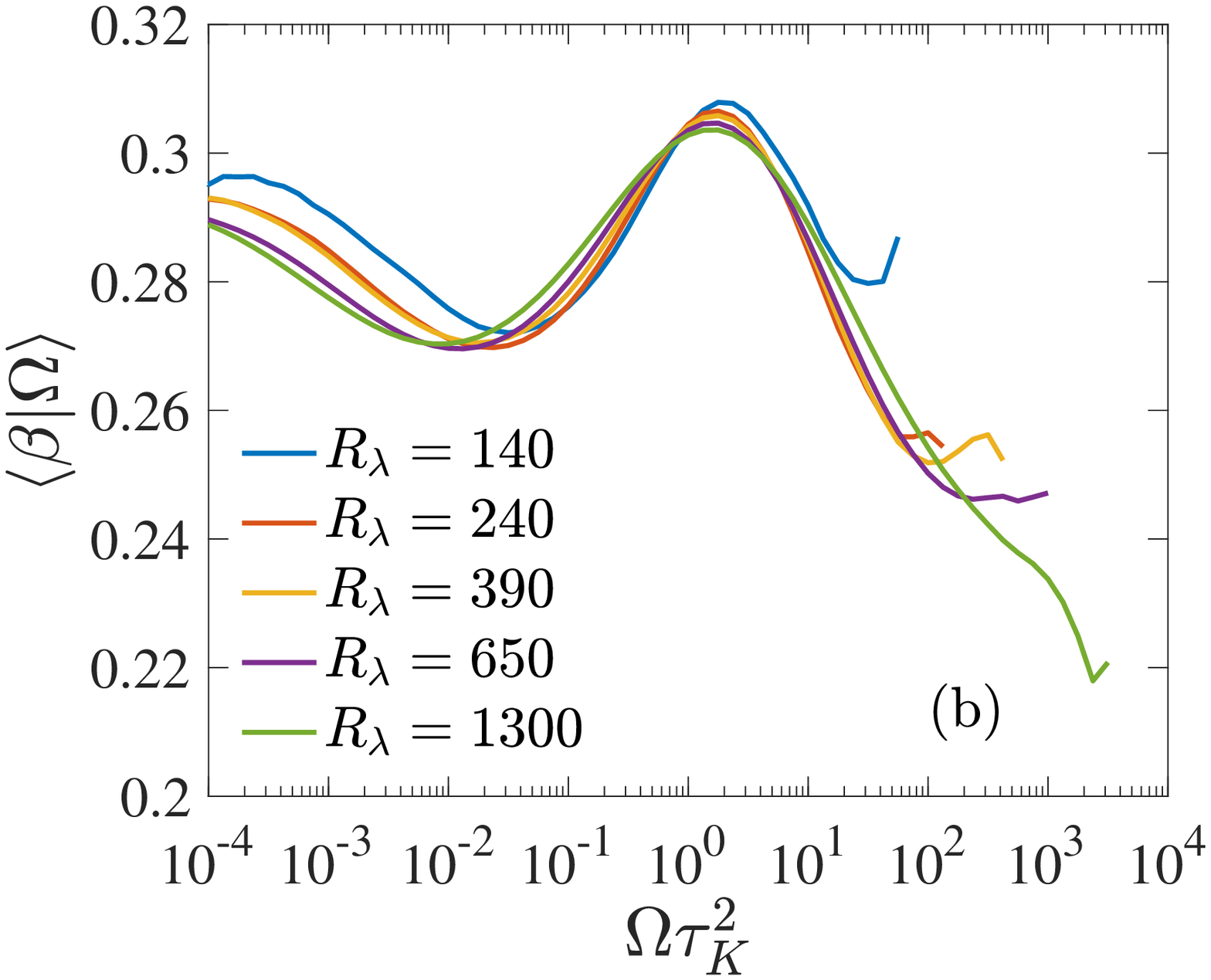}  \
\includegraphics[width=0.31\textwidth]{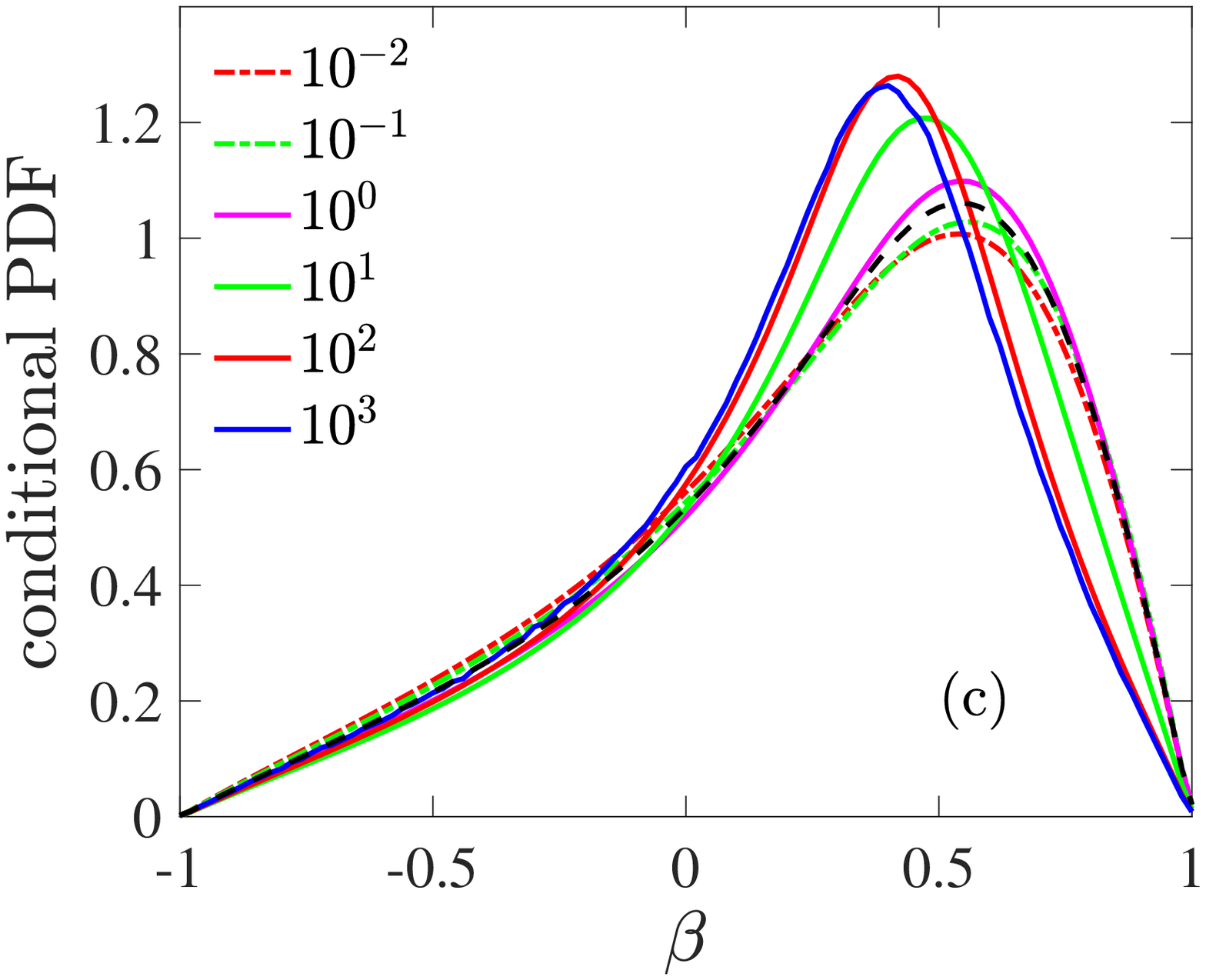} 
\caption{
Conditional expectations of (a) first two eigenvalues of strain tensor, 
non-dimensionalized by $\tau_K$, and
(b) $\beta$ as defined by Eq.~\eqref{eq:beta}, for various $\re$.
The dashed lines in (a) correspond to a power law $\Omega^{1/2}$.
The endmost bins for $\beta$ (at comparatively large conditioning values for each
$\re$) are not converged properly, since calculating $\beta$ involves
cancellations between first and third eigenvalues (which are much larger 
than the intermediate eigenvalue).
(c) Conditional PDFs of $\beta$ at $\re=1300$, for same conditioning values 
of $\Omega\tau_K^2$ as used in Fig.~\ref{fig:align}. The black dashed line
represents the marginal PDF, as shown earlier in Fig.~\ref{fig:g_align}b.
}
\label{fig:lam12}
\end{center}
\end{figure}

The conditional relation between strain and vorticity is further
analyzed by considering the conditional expectations of eigenvalues, i.e.,
$\langle \lambda_i |\Omega \rangle$.
Figure~\ref{fig:lam12}a shows the conditional expectations 
for $i=1$ and $2$. 
The properties of the third eigenvalue can be easily inferred
from the relation $\lambda_3 = - ( \lambda_1 + \lambda_2)$. 
As evident from Fig.~\ref{fig:lam12}a,
we find that the conditional expectation of  
$\lambda_2$ is always positive, throughout the entire range of 
$\Omega$ and smaller than that of $\lambda_1$, approximately
by a constant factor $\approx 8$ 
(note that $\lambda_1 $ is by construction always positive).
Furthermore, both eigenvalues show the same qualitative behavior,
matching that of $\langle \Sigma |\Omega \rangle$.
This is not surprising since $\Sigma$ can be written as
$\Sigma = 2(\lambda_1^2 + \lambda_2^2 + \lambda_3^2)$.
Evidently, the contribution to $\Sigma$ is dominated by
$\lambda_1$ and $\lambda_3$ since their magnitudes are
significantly larger than that of $\lambda_2$.
Consequently, one can expect $\lambda_1$ (and -$\lambda_3$)
to scale as $\Sigma^{1/2}$ (even when conditioned
on $\Omega$). Although not explicitly shown, 
we indeed find this to be the case, with
$\langle \lambda_1  |\Omega \rangle $
and $\langle -\lambda_3  |\Omega \rangle $  approximately scaling
as $\Omega^{\gamma/2}$.

Using similar arguments as above,
one could possibly expect
$\langle \lambda_2 |\Omega \rangle$ to also scale as 
$\Omega^{\gamma/2}$.
However, this is not the case and the approximate
power law we find corresponds to an exponent smaller than $\gamma/2$. 
This can be readily
observed by considering either the ratio $\lambda_2/\lambda_1$
or the parameter $\beta$ defined in Eq.~\eqref{eq:beta},
which is bounded between $-1$ and $1$. 
The conditional expectation $\langle \beta|\Omega\rangle$ is
shown in Fig.~\ref{fig:lam12}b. 
For $\Omega \tau_K^2\lesssim 1$, $\beta$ does not vary much, 
but for larger $\Omega$, $\beta$ appears to slowly
decrease, almost independently of $\re$. 
Thus, it appears reasonable to assume
$\langle \lambda_2|\Omega \rangle \sim \Omega^{\gamma/2 - \delta}$.
Our data suggest $\delta \approx  0.05$
(with the coefficient of determination exceeding $99\%$).

To further investigate the behavior of the intermediate eigenvalue,
we show the conditional PDFs of $\beta$ in Fig.~\ref{fig:lam12}c
for $\re=1300$ (with the marginal PDF in black dashed line).
Consistent with $\langle \beta|\Omega\rangle$, 
the PDFs show very minor variations for small
$\Omega\tau_K^2$. For larger $\Omega \tau_K^2$,
a slightly stronger variation can be seen, with the PDFs
peaking at slightly smaller positive values of $\beta$.
Interestingly, for all conditioning values, there
is virtually  no change in
PDFs for negative value of $\beta$.
This suggests that the likelihood of $\lambda_2$ to be negative
(or positive) stays the same, with just positive values becoming
comparatively smaller in magnitude at large $\Omega$. 
This suggests that for intense vorticity events,
there is a somewhat stronger cancellation between
 $\lambda_1 $ and $\lambda_3$. This is consistent with the 
notion that the regions where vorticity is most intense are essentially
two-dimensional~\cite{Jimenez:1992}, with the largest and the smallest 
eigenvalues of strain almost canceling each other, and 
the value of the intermediate eigenvalue being much smaller.
This is also corroborated by the observation that
the exponent of the
power law for $\langle \lambda_2 | \Omega \rangle$ is 
smaller than for $\langle \lambda_1 | \Omega \rangle$
(and  $\langle -\lambda_3 | \Omega \rangle$).

\subsection{Enstrophy production}
\label{subsec:enstr_budget}

\begin{figure}
\begin{center}
\includegraphics[width=0.47\textwidth]{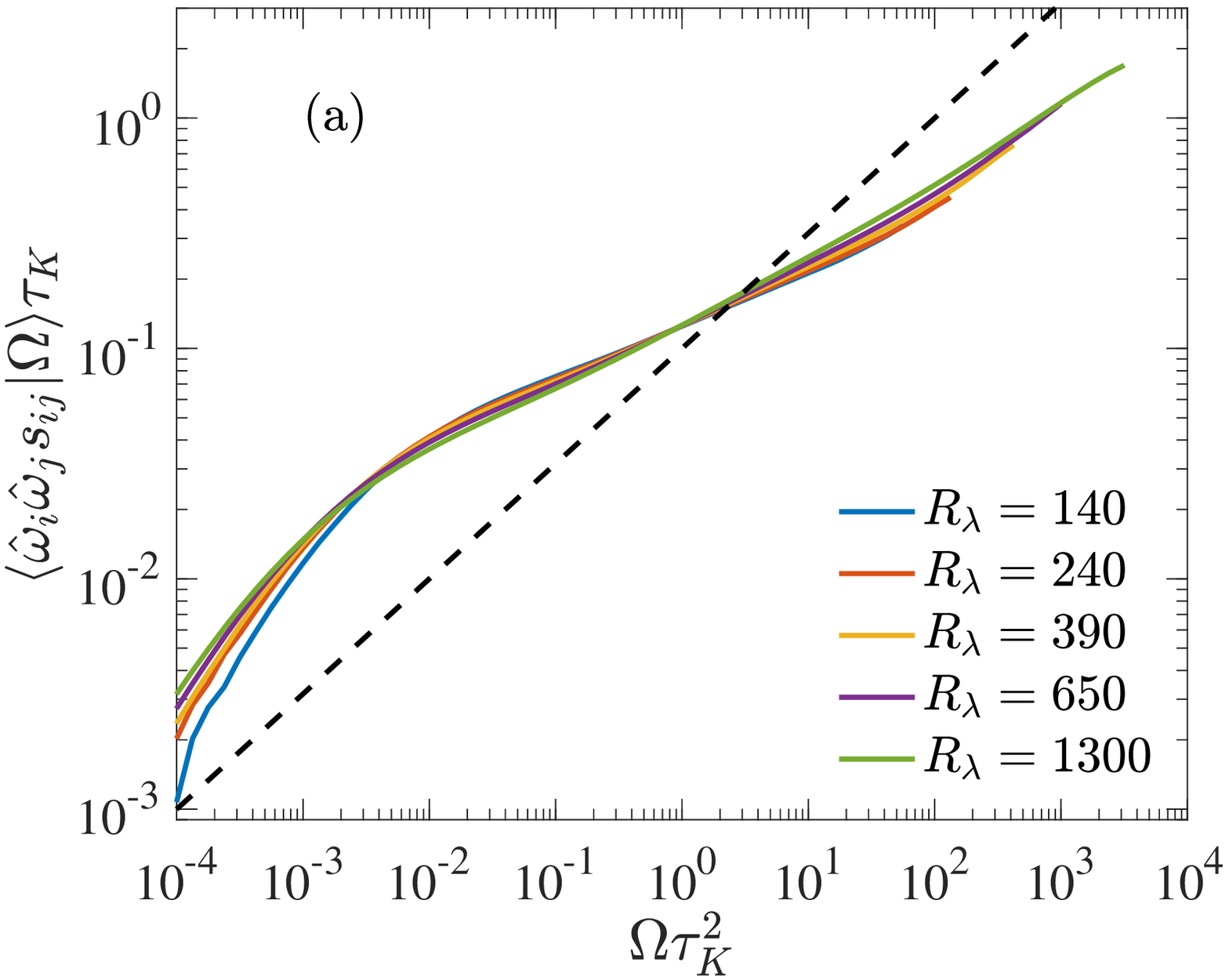} \ \ \ \ 
\includegraphics[width=0.47\textwidth]{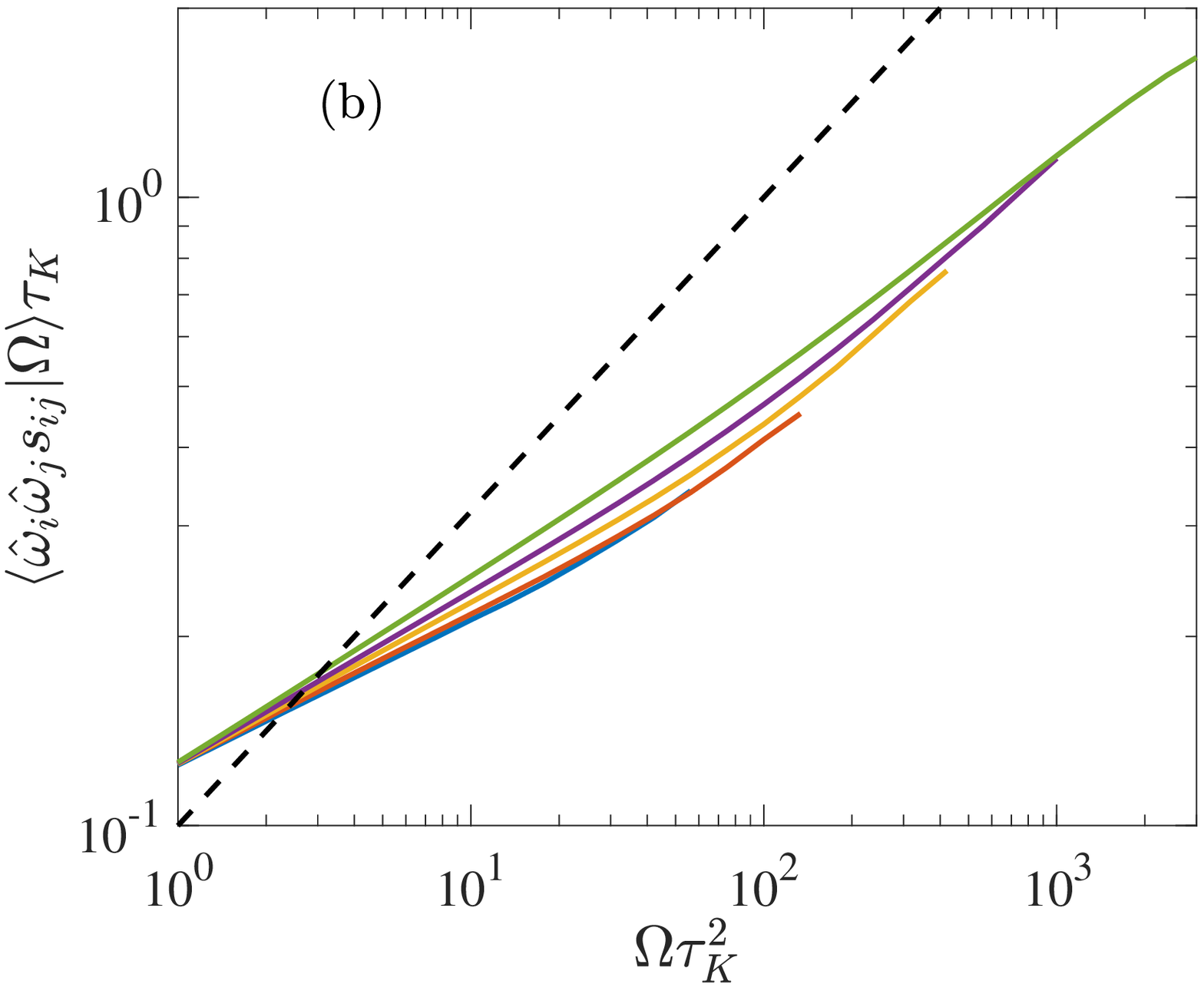} \\
\includegraphics[width=0.47\textwidth]{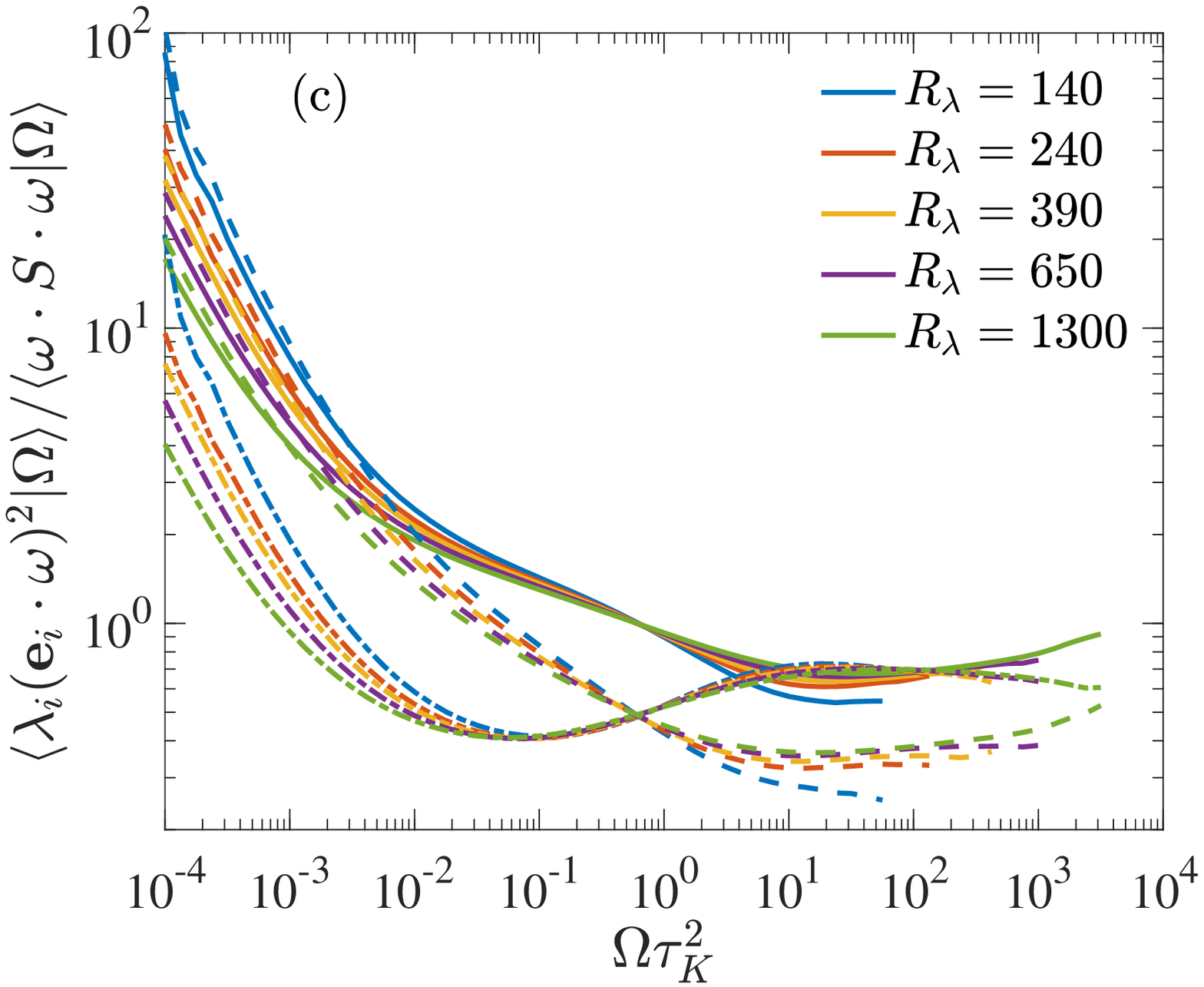}  \ \ \ \ 
\includegraphics[width=0.47\textwidth]{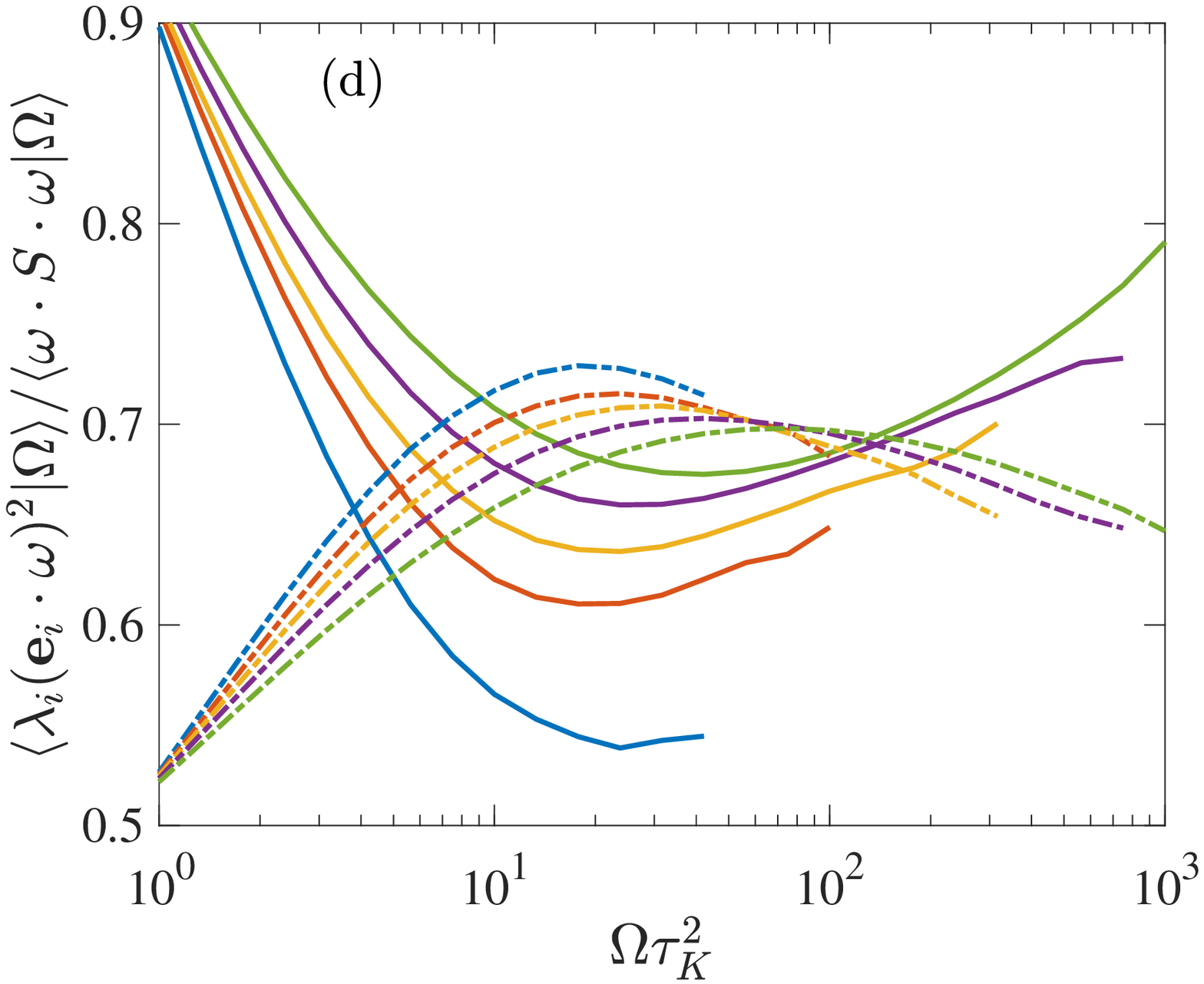}  
\caption{
(a) Conditional expectation of the net enstrophy
production term, written as an effective strain
and non-dimensionalized by $\tau_K$, at various $\re$.
The dashed line represents a power law of $\Omega^{1/2}$.
(b) Zoomed in version of (a).
(c) The relative contributions from each eigendirection
to the net production of enstrophy at various $\re$.
Solid, dashed-dotted and dashed lines for
$i=1$, $2$ and $3$ respectively. For $i=3$, the negative
of the quantity is shown.
(d) Zoomed in version of (c), only showing $i=1$ and $2$.
}
\label{fig:wsw}
\end{center}
\end{figure}

We now investigate the net enstrophy production resulting from the
correlations between strain and vorticity discussed earlier.
Figure~\ref{fig:wsw}a shows the conditional average 
$\langle \hat{\omega}_i  \hat{\omega}_j s_{ij} | \Omega \rangle$
(made dimensionless by multiplying by $\tau_K$),
which simply equals
$ \langle \omega_i \omega_j s_{ij}| \Omega \rangle/\Omega$,
since $\Omega$ is the conditioning variable.
The conditional average 
$\langle \hat{\omega}_i \hat{\omega}_j s_{ij} | \Omega \rangle$
can also be interpreted as the {\em effective} strain 
acting on vorticity, which ultimately leads 
to production of enstrophy.
The dashed line shown in 
Fig.~\ref{fig:wsw}a represents a $\Omega^{1/2}$ dependence, 
which would correspond to strain as strong as the local vorticity.
Figure~\ref{fig:wsw}a shows that for very small values of $\Omega \tau_K^2$, 
$\langle \hat{\omega}_i  \hat{\omega}_j s_{ij} | \Omega \rangle$
grows faster than $\propto \Omega^{1/2}$ -- despite
earlier result in Fig.~\ref{fig:sigma} showing that the magnitude
of strain conditioned on vorticity is mostly constant for small $\Omega$.
This can be understood by realizing first that 
for $\Omega \to 0$,
the conditional production term 
also goes to zero, since the strain and vorticity are completely decorrelated.
However, as $\Omega$ increases, while the strain and vorticity magnitude are still
decorrelated, there is some weak preferential alignment between vorticity 
and first two eigenvectors of strain (as noticed in Fig.~\ref{fig:align}).
This incipient alignment is in precise agreement with the strong
initial growth of production term.

On the other hand, for large $\Omega$,
the alignments are approximately in an asymptotic state,
whereas $\langle \Sigma | \Omega \rangle$ is weaker than $\Omega^1$ --
suggesting a similar behavior 
$\langle \hat{\omega}_i \hat{\omega}_j s_{ij}| \Omega \rangle$.
The results shown in 
Fig.~\ref{fig:wsw}a are consistent with this expectation.
To better show the dependence on $\re$, Fig.~\ref{fig:wsw}b
zooms on the domain $\Omega \tau_K^2 \gtrsim 1$.
The $\re$ dependence appears to be qualitatively similar to that
also observed for strain and its eigenvalues in Section~\ref{subsec:strain}.
However, given the different power law behaviors of 
$\langle \lambda_1 |\Omega\rangle $ and 
$\langle \lambda_2 |\Omega\rangle $,
it is reasonable to model the curves in
Fig.~\ref{fig:wsw}b as a sum of these two power laws:
\begin{align}
\langle \hat{\omega}_i \hat{\omega}_j s_{ij}| \Omega \rangle \approx
c_1 (\Omega \tau_K^2)^{\gamma/2}  + c_2 (\Omega \tau_K^2)^{\gamma/2-\delta} 
\label{eq:fit_str}
\end{align}
Note, such a functional form assumes that the conditional alignments
are approximately constant with $\Omega$ (for $\Omega\tau_K^2 \gtrsim1$),
which is justified from the observation in Fig.~\ref{fig:align}d.
In earlier works \cite{wilczek09,johnson16}, this dependence was modeled 
as a single power law, leading to some inconsistencies, 
until an ad hoc tuning of the model was performed.  
Our data suggest $c_1\approx 0.088$ and $c_2\approx c_1/20$
(with the coefficient of determination exceeding $99\%$).

Further insight into enstrophy production is provided by
Fig.~\ref{fig:wsw}c, 
which shows the fractional contribution in each eigendirection
to the total production (based on Eq.~\eqref{eq:weiw}).
Since the contribution corresponding to $\lambda_3$ is always negative,
we plot its absolute magnitude, for explicit comparison with the 
contributions from other eigenvalues on log-log scales.
Two very different behaviors corresponding to very weak and 
very large events respectively are observed. For weak events,
the individual relative contributions are all significantly larger 
than unity,  implying a strong cancellations to produce
the net production of enstrophy (we recall that the individual 
contributions sum up to unity). 
This is also consistent with the observation above that the lack 
of alignment
between vorticity and the eigenvalues of strain at small values of 
$\Omega \tau_K^2$ results in a sum of the contributions 
of the three eigenvectors $\approx 0$ i.e. much weaker than 
each individual term. 
A consequence of the very weak alignment of vorticity with 
any of the three strain eigenvectors (see Fig.~\ref{fig:align}d) 
is that these individual contributions
can be estimated to be in the ratio of 
the magnitude of eigenvalues, which indeed appears to be the case.
However, as the value of $\Omega \tau_K^2$ increases, 
vorticity aligns 
preferentially with some particular eigenvectors of strain, and the
individual contributions to the net productions change accordingly.

To better understand the $\re$ dependence, 
Fig.~\ref{fig:wsw}d shows a zoomed in version of Fig.~\ref{fig:wsw}c,
focusing only on the first and second eigendirections. 
The relative contribution to enstrophy
production from the first 
eigendirection start decreasing in magnitude when $\Omega \tau_K^2$ 
increases, 
whereas the positive contribution from the
intermediate eigendirection starts increasing. 
For extreme events, the contributions from the first and the second 
eigendirections appear to be comparable (whereas
that from the third direction is significantly smaller 
in magnitude).
It turns out that the contribution from the second direction
is slightly larger than that from first for moderate events.
However, this difference clearly becomes smaller as $\re$ increases.
On the other hand, for most intense events,
the dominant contribution always comes 
from the first direction, despite the 
strong tendency of vorticity to align with intermediate
eigenvector. In addition, the contribution from
first direction also appears to slowly increase
(with the conditioning value), whereas the contribution for
the second direction appears to slowly decrease.

\subsection{Viscous destruction of enstrophy}

\begin{figure}
\begin{center}
\includegraphics[width=0.47\textwidth]{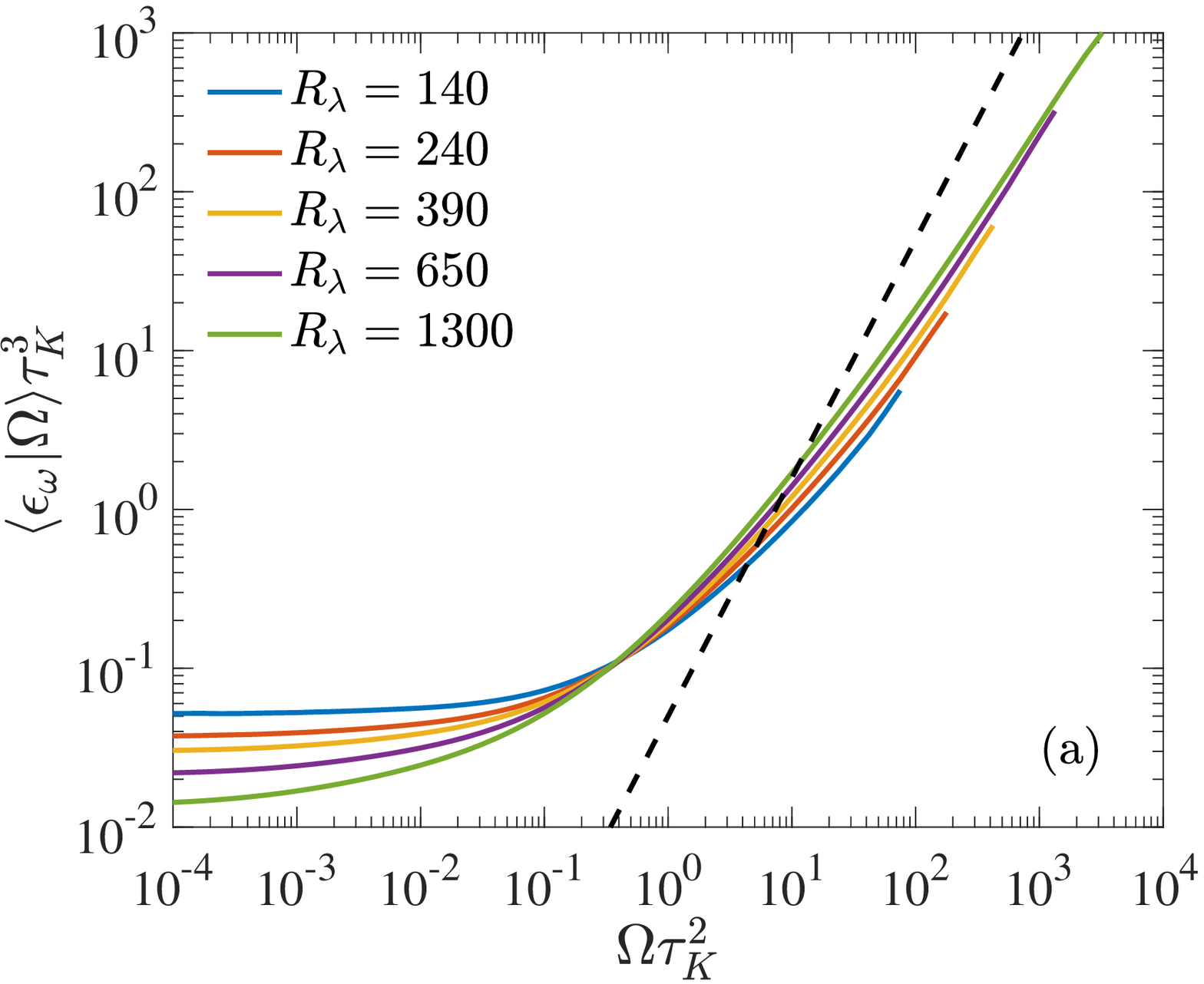} \ \ \ \
\includegraphics[width=0.47\textwidth]{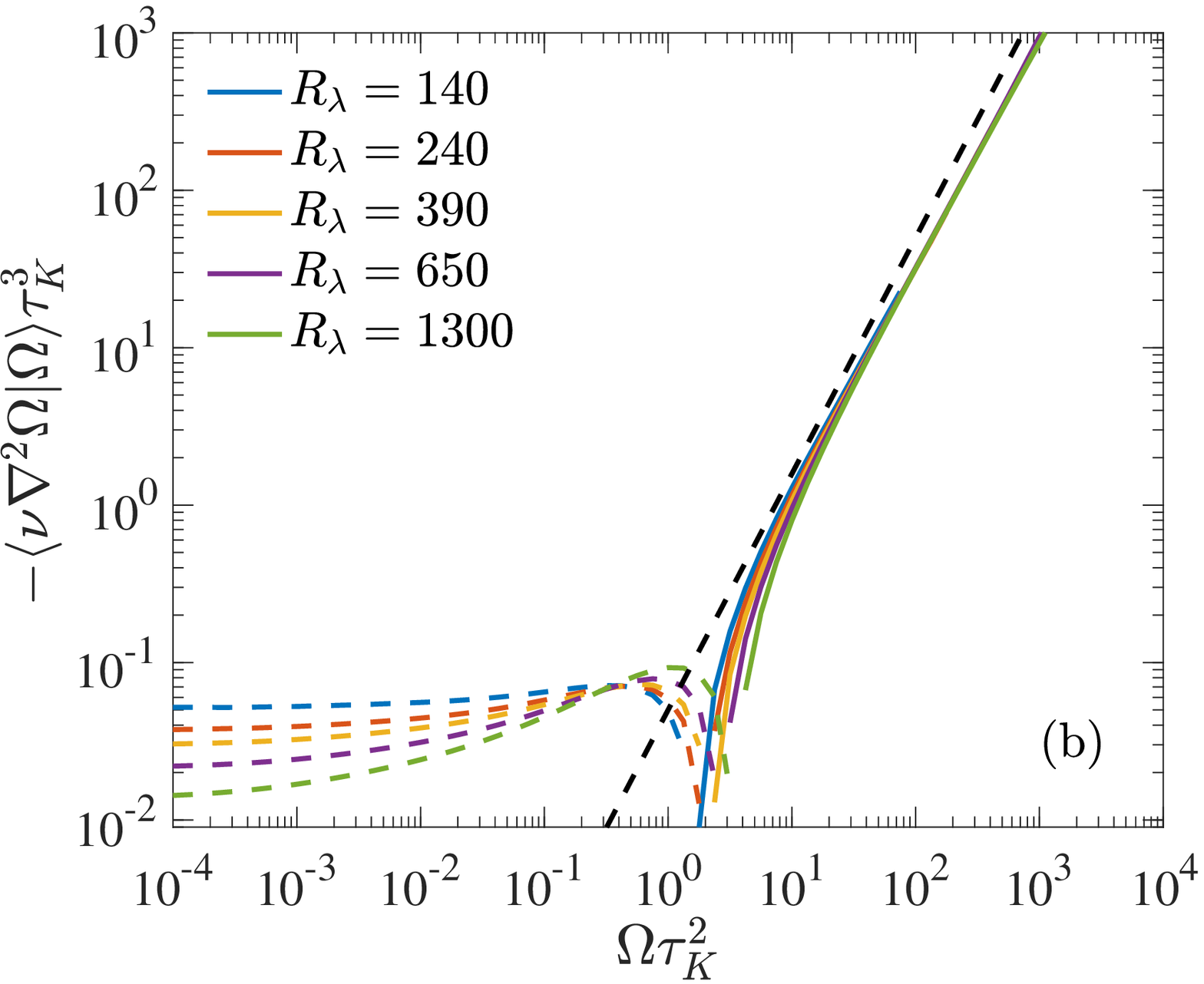}
\caption{
Conditional expectations of
(a) the viscous dissipation of enstrophy, and 
(b) viscous diffusion of enstrophy -- 
as shown in Eq.~\eqref{eq:enstr} -- at various $\re$. 
The dashed lines in both represent a power law of $\Omega^{3/2}$.
In (b), the portion in dashed lines
corresponds to negative contribution.
}
\label{fig:wsw2}
\end{center}
\end{figure}

While vortex stretching on an average leads to enstrophy production,
the viscous terms
lead to enstrophy dissipation,
with their effects balancing each other
in the (statistically) stationary state. 
Of the two viscous terms in Eq.~\eqref{eq:enstr},
the term  
$\epsilon_w = 2 \nu \frac{\partial \omega_i }{\partial x_j} 
\frac{\partial \omega_i }{\partial x_j} $  contributes 
to enstrophy dissipation, whereas the other term
$\nu\nabla^2\Omega$ is on average zero (because of homogeneity) 
and only leads to
spatial redistribution of enstrophy.
However, when conditioning on $\Omega$, the second term
may contribute to the enstrophy budget differently,
depending on the strength of vorticity \cite{holzner10}. 
Figures~\ref{fig:wsw2}a and b show the conditional averages of $\epsilon_w$ 
and $-\nu \nabla^2 \Omega$ respectively, appropriately 
non-dimensionalized by $\tau_K^3$. 
While $\epsilon_\omega$ is always positive, 
the requirement  
$\langle \nabla^2 \Omega \rangle = 0$ imposes that
the quantity $-\nu \langle \nabla^2 \Omega | \Omega \rangle$
cannot be positive definite.
For this reason and also since
Fig.~\ref{fig:wsw2}b utilizes log scales, 
positive values of $-\nu \langle \nabla^2 \Omega | \Omega \rangle$, 
which correspond to destruction of enstrophy,
are represented by solid lines, whereas 
negative values corresponding to production of enstrophy,
are shown in dashed lines.

As performed in previous subsections, we can demarcate the results
based on the strength of vorticity.
As evident in Fig.~\ref{fig:wsw2}, both terms are roughly constant 
at small $\Omega$ conditioning values
and of virtually identical magnitude, but of opposite signs.
This suggests that the overall contribution of viscous terms 
at small $\Omega$ values is negligible.
For $\Omega\tau_K^2 \gtrsim 1$,
both terms are positive and strongly increase with $\Omega$.
The latter is consistent with the notion that the viscous diffusion term
overall acts to deplete enstrophy in regions of high enstrophy
and thereafter diffuse it to regions of weaker enstrophy
(resulting in no net contribution to average enstrophy budget).
Similar to Fig.~\ref{fig:sigma}, the curves for 
$\langle\epsilon_\omega|\Omega\rangle$ appear to get steeper with increasing $\re$.
On the other hand, the curves for
$-\nu \langle \nabla^2 \Omega | \Omega \rangle$ remarkably do not appear
to change slope with increasing $\re$
(but are slightly shifted, with the zero-crossing point
weakly dependent on $\re$).

Since both the viscous terms are
dimensionally equivalent to the enstrophy production term,
which in turn is equivalent to third moment of vorticity,
we compare the curves in Fig.~\ref{fig:wsw2}a and b with dashed line
representing $\Omega^{3/2}$.
Similar to enstrophy production term (in Fig.~\ref{fig:wsw}a and b),
the growth of $\langle \epsilon_w | \Omega \rangle$ is slower
than $\Omega^{3/2}$, though similar trends with $\re$ are observed in both cases
with the curves becoming steeper and slightly closer to $\Omega^{3/2}$.
On the other hand, for $-\nu \langle \nabla^2 \Omega | \Omega \rangle$,
all the curves appear to conform to $\Omega^{3/2}$ scaling.
Interestingly, we observe that while both conditional viscous terms
are of comparable magnitude, the diffusion
term dominates at large $\Omega$ values,
suggesting that destruction of enstrophy is dominated by the diffusion
term for the most extreme events.

To interpret the results of Fig.~\ref{fig:wsw2},
we recall that $\nu \nabla^2 \Omega$ describes 
a diffusion process, transferring enstrophy from
regions where it is large 
to regions where it is  smaller.
Figure~\ref{fig:wsw2} allows us to distinguish two ranges,
for $ \Omega \tau_K^2 \lesssim 1$, and for $\Omega \tau_K^2 \gtrsim 1$. 
The ostensibly simple dependence of 
$\langle \nu \nabla^2 \Omega | \Omega \rangle \propto \Omega^{3/2}$ 
for $\Omega \tau_K^2 \gtrsim 1$ can possibly be
explained as follows.
Using the observation that the largest vorticity 
fluctuations correspond to narrow weakly curved vortex tubes, 
we can approximately write 
$\nu \nabla^2 \Omega \sim \nu \Omega/\ell^2$,
where $\ell$ is the characteristic radius of the vortex tube.
Thereafter, taking $\nu \Omega/\ell^2 \sim \Omega^{3/2}$
implies that 
$\Omega^{1/2}\ell^2/\nu \sim 1$.
Thus, the vortices
that contribute to viscous diffusion have  a size such that the local 
Reynolds number is of order unity. 
Interestingly, a similar argument is often utilized in defining
a local fluctuating dissipation length scale, akin to a multifractal 
description \cite{Frisch95}. However, note the current argument stems from considering
the dissipation of vorticity, and is conceptually  different
than the other one, which relies on a local energy balance 
(and does not adequately describe the intense vortices \cite{Jimenez93,BPBY2019}).

\section{Conclusions}
\label{sec:concl}

The vortex stretching mechanism, resulting from
non-linear coupling of vorticity $\omega_i$
and strain-rate tensor $s_{ij}$, 
plays a central role in the formation of extreme events and small-scales 
in turbulent flows. In this paper,
we have systematically investigated the statistical
correlations underlying the vortex stretching mechanism
in direct numerical simulations of 
stationary isotropic turbulence across a wide range of 
Reynolds numbers ($140 \le \re \le 1300$).
As commonly done, the correlations are studied in the eigenframe 
of the strain-rate tensor -- for which, by definition, 
the first and third eigenvalues are
positive and negative respectively. 
In this context, it is well known that vorticity 
preferentially aligns with the eigenvector of strain corresponding to the intermediate 
eigenvalue, $\lambda_2$, which is on the average positive
and thus leads to net production of enstrophy (vorticity-squared) 
This behavior is very robust across all $\re$,
as shown in Fig.~\ref{fig:g_align}.
However, as suggested before \cite{Tsi2009},
this does not necessarily imply that the dominant contribution
to production of enstrophy also arises from the intermediate eigendirection.
We confirm this and show that the dominant 
contribution arises
from the first eigendirection, since the corresponding eigenvalue is significantly
larger than the intermediate eigenvalue. However, the contribution from  the
second (intermediate) eigendirection is also comparable.

In order to understand the formation and structure of extreme
events \cite{BPBY2019}, we investigated the 
coupling between vorticity and strain tensor 
by extracting statistics conditioned on the enstrophy 
$\Omega = \omega_i \omega_i$.
The alignment between vorticity and the eigenvector corresponding to the
intermediate eigenvalue of strain is found to be much stronger 
than the averaged alignment shown in Fig.~\ref{fig:g_align}a
when $\Omega \tau_K^2 \gg 1$, where $\tau_K$ is the Kolmogorov
time scale defined as $\tau_K^2 = 1/\langle \Omega \rangle$.
This trend is consistent with visualizations of the most intense
vorticity structures in turbulent flows, which turn out to be almost 
straight vortex tubes; and also with many attempts to follow numerically the 
formation  of very strong velocity gradients in the Euler equations, which 
also revealed a strong tendency to form quasi two-dimensional structures, where
vorticity can only be aligned with the intermediate strain eigenvector. 
Due to this enhanced alignment, we find that the contribution
of second eigendirection is larger than that from first for 
moderately strong vorticity events,
however the difference becomes smaller as $\re$ increases. 
In contrast, for the most extreme events ($\Omega\tau_K^2 \gtrsim 100$), 
the contribution from first eigendirection is always larger
than that from second, albeit they are still comparable in magnitude.

Interestingly, we find that the conditional averages of strain magnitude
(given by $\Sigma=2s_{ij}s_{ij}$) 
and the eigenvalues of strain,
are mostly constant for $\Omega\tau_K^2 < 1$,
and approximately vary 
power laws for $\Omega\tau_K^2 >1$.   The exponent $\gamma$,
which describes the power law for $\Sigma$, i.e.,  
$\langle \Sigma | \Omega \rangle \tau_K^2 \sim  (\Omega \tau_K^2)^{\gamma}$, 
also describes the dependence of the first and third eigenvalue,
where $\gamma$ is a function of $\re$,
in accordance with \cite{BPBY2019}. However, a minor correction, independent
of $\re$, is necessary for the intermediate eigenvalue.
Combining the results, the enstrophy production can also be expressed by 
a simple functional form, involving the sum of two power laws, 
see Eq.~\eqref{eq:fit_str}. 
In contrast, an analysis of the viscous terms responsible for destruction
of enstrophy reveals that the primary contribution for large enstrophy comes
from viscous diffusion of enstrophy ($\nu \nabla^2 \Omega$), 
rather than viscous dissipation ($\nu ||\nabla \ww ||^2$). 
Whereas for small enstrophy the diffusion term actually leads to 
enstrophy production (since ultimately  this term does not
contribute to the overall enstrophy budget).
We highlight the importance of these results
in light of turbulence modeling (albeit an explicit attempt is
left for future work).

In discussing our results, we have used several times 
the remark that the most intense vortex structures are shaped as weakly curved
vortex tubes~\cite{Jimenez93,Ishihara07,BPBY2019}, which, as noticed by
\cite{Jimenez:1992}, may provide a kinematic explanation for the observed
alignment between vorticity and the eigenvectors of strain. Our results,
obtained by conditioning the statistics on 
enstrophy, appear to corroborate this picture. 
However, some of the other results uncovered in this work,
such as the conditional expectation of $\beta$, 
cannot be simply explained by assuming that the regions 
with most intense vorticity and strain are due to weakly curved 
Burgers vortices. This suggests existence of a more
complicated structure of the vortex tubes, involving also axial 
velocity~\cite{Choi:09}. 
In this light, it is also important to note that the analysis 
performed here purely relies on utilizing single-point correlations.
However, vorticity and strain are non-locally related to each other
through the Biot-Savart relation \cite{ham_pof08,buaria_nc}.
Thus, a complete picture of vortex stretching can only emerge
by also characterizing the non-local effects.
In this regard, it is also important to recognize the role
of pressure Hessian in affecting the dynamics of strain rate tensor
\cite{nomura:1998}. 
In this regard, a careful analysis of the non-local effects 
is ongoing and will be reported separately.

We conclude by noticing that, although we have observed over the range 
of Reynolds numbers considered ($140 \le \re \le 1300$)
a systematic dependence on $\re$ of the statistical properties of the 
velocity gradient tensor conditioned on enstrophy or strain, the intriguing 
question remains as to whether the quantities studied here tend to a
limiting form when $\re \rightarrow \infty$.

\section*{Acknowledgments}

We gratefully acknowledge the Gauss Centre for Supercomputing e.V.
(www.gauss-centre.eu) for funding this project by providing computing time on the
GCS supercomputer JUQUEEN and JUWELS at J\"ulich Supercomputing Centre (JSC),
where the simulations reported in this paper were performed.
We acknowledge support from
the  Max Planck Society.
We also thank P. K. Yeung for sustained collaboration
and partial support
under the Blue Waters computing project
at the University of Illinois Urbana-Champaign.

\appendix

\section{Effect of spatial resolution}
\label{app:resol}

\begin{figure}
\begin{center}
\includegraphics[width=0.47\textwidth]{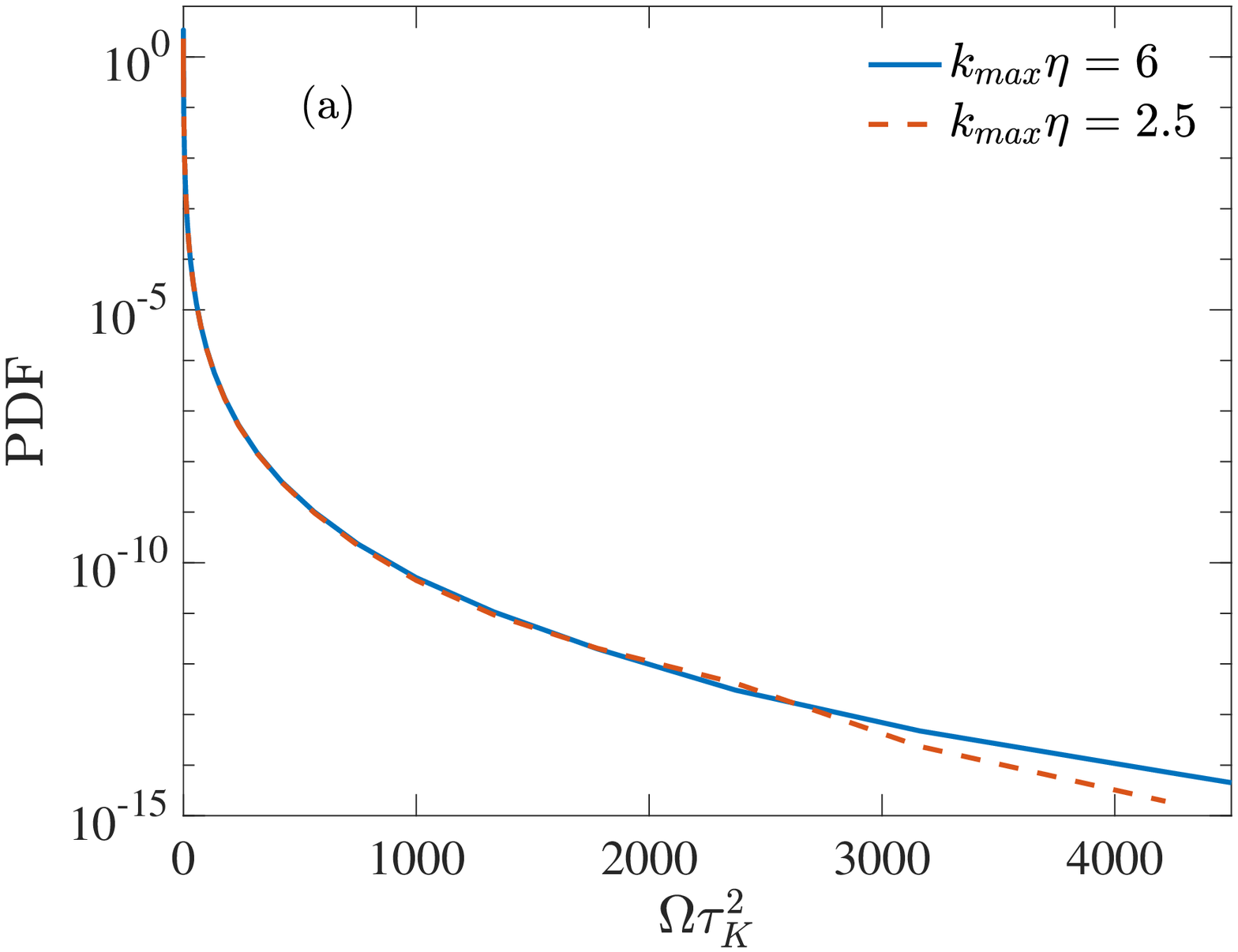} \ \ \ \  
\includegraphics[width=0.47\textwidth]{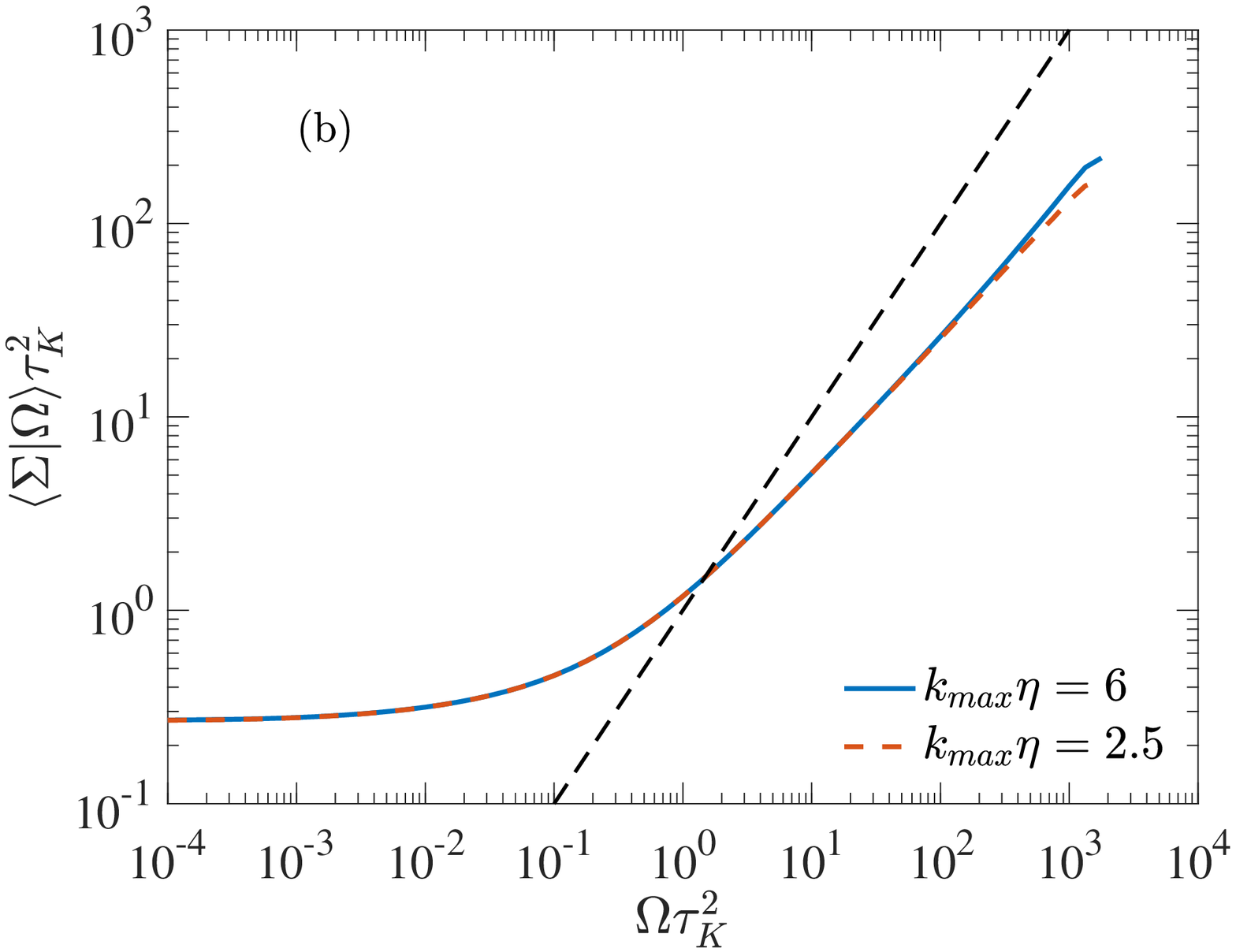} 
\caption{
(a) Probability density function (PDF) of enstrophy at $\re=650$,
corresponding to two different small-scale resolutions.
(b) Expectation of strain conditioned on enstrophy for same cases.
The dashed line represents a slope of $1$.
The effect of resolution is negligible, only visible for the most extreme 
events, is overall negligible.
}
\label{fig:res}
\end{center}
\end{figure}

In this appendix, we briefly address the concerns pertaining 
to small-scale resolution. 
As listed earlier, all the runs up to $\re=650$ in this work have a spatial
resolution of $k_{max}\eta=6$, which is more than adequate
to resolve the extreme events \cite{BPBY2019}.
In fact in \cite{BPBY2019}, it was suggested
that the spatial resolution necessary to accurately resolve the extreme events
is given as
\begin{align}
k_{max}\eta \approx 3 (\re/\re^{ref})^\alpha \ ,
\label{eq:resol}
\end{align}
where $\re^{ref}=300$ and $\alpha=0.275\pm0.025$.
This formula suggests that at $\re=1300$, one would ideally 
require $k_{max}\eta \approx 4.5$. 
However, the run at $\re=1300$ corresponds to $k_{max}\eta=3$. 
Nevertheless, we show below that conditional statistics studied here
are not very sensitive to the resolution~\cite{yeung2006}.
Since we do not have data at higher resolution for $\re=1300$,
we consider the $\re=650$ data at a smaller resolution.
Utilizing Eq.~\eqref{eq:resol}, a resolution of $k_{max}\eta=3$
(instead of $4.5$) at $\re=1300$, would correspond
to $k_{max}\eta\approx2.5$ at $\re=1300$, if both are 
under resolved in a similar fashion
(note for $\re=650$, the correct resolution would be
$k_{max}\eta\approx3.8$).
We present the results in Fig.~\ref{fig:res}, comparing
the PDF of enstrophy in (a), and the conditional
expectation $\langle \Sigma|\Omega \rangle$ in (b). 
Other conditional statistics discussed in this study behave similarly to 
the result in Fig.~\ref{fig:res}b.


%

\end{document}